\documentclass[10pt,journal,compsoc]{IEEEtran}

\ifCLASSOPTIONcompsoc
  \usepackage[nocompress]{cite}
\else
  \usepackage{cite}
\fi
\usepackage{amsmath}
\usepackage{xcolor}
\usepackage{framed}
\usepackage{multicol}
\usepackage{comment} 
\usepackage{paralist}
\usepackage{url}
\usepackage{amsthm}
\usepackage[inline]{enumitem}
\usepackage{siunitx}
\usepackage{subcaption}
\usepackage{multirow}
\usepackage{tcolorbox}
\usepackage{booktabs}

\theoremstyle{definition}
\newtheorem{definition}{Definition}

\usepackage[linesnumbered, ruled, vlined]{algorithm2e}
\SetAlFnt{\small}
\SetAlCapFnt{\small}
\SetAlCapNameFnt{\small}

\usepackage{tipa}

\usepackage{xspace}
\newcommand{\MLCSHE}{MLCSHE\xspace}
\newcommand{\app}{MLCSHE\xspace}

\usepackage{newtxmath}

\usepackage[numbers]{natbib}

\ifCLASSINFOpdf
\else
\fi

\usepackage{hyperref}
\hypersetup{
colorlinks,
allcolors=black,
urlcolor={black},
pdftitle={Identifying the Hazard Boundary of 
ML-enabled Autonomous Systems Using Cooperative Co-Evolutionary Search},
pdfsubject={ML System Safety},
pdfauthor={Sepehr Sharifi, Donghwan Shin, Lionel C. Briand, Nathan Aschbacher},
pdfkeywords={ML-enabled Autonomous System, Hazard Boundary, System Safety Monitoring, Cooperative Co-Evolutionary Search}
}

\begin{document}
\title{Identifying the Hazard Boundary of\\
ML-enabled Autonomous Systems Using Cooperative Co-Evolutionary Search}

\author{Sepehr~Sharifi
        ,
        Donghwan~Shin
        ,~\IEEEmembership{Member,~IEEE,}
        Lionel~C.~Briand
        ,~\IEEEmembership{Fellow,~IEEE,}
        and Nathan Aschbacher%
\IEEEcompsocitemizethanks{\IEEEcompsocthanksitem S. Sharifi and L. Briand are with the Department
of Electrical and Computer Engineering, University of Ottawa, Ottawa,
Ontario, Canada, K1N 5N6. L. Briand has also a faculty appointment with the SnT Centre at the University of Luxembourg, Luxembourg. \protect\\
E-mail: \{s.sharifi, lbriand\}@uottawa.ca
\IEEEcompsocthanksitem D. Shin is with Department of Computer Science, University of Sheffield, Sheffield, United Kingdom, S1 4DP
\protect\\
E-mail: d.shin@sheffield.ac.uk
\IEEEcompsocthanksitem N. Aschbacher is with Auxon Corporation, Portland, Oregon, United States and its subsidiary Auxon Technologies, Ottawa, Ontario, Canada
\protect\\
E-mail: nathan@auxon.io}
} %

\IEEEtitleabstractindextext{%
\begin{abstract}
In Machine Learning (ML)-enabled autonomous systems (MLASs), it is essential to identify the \emph{hazard boundary} of ML Components (MLCs) in the MLAS under analysis. Given that such boundary captures the conditions in terms of MLC behavior and system context that can lead to hazards,  it can then be used to, for example, build a safety monitor that can take any predefined fallback mechanisms at runtime when reaching the hazard boundary.
However, determining such \emph{hazard boundary} for an ML component is challenging.
This is due to the problem space combining system contexts (i.e., scenarios) and MLC behaviors (i.e., inputs and outputs) being far too large for exhaustive exploration and even to handle using conventional metaheuristics, such as genetic algorithms. Additionally, the high computational cost of simulations required to determine any MLAS safety violations makes the problem even more challenging. 
Furthermore, it is unrealistic to consider a region in the problem space deterministically safe or unsafe due to the uncontrollable parameters in simulations and the non-linear behaviors of ML models (e.g., deep neural networks) in the MLAS under analysis.
To address the challenges, we propose \MLCSHE (ML Component Safety Hazard Envelope), a novel method based on a Cooperative Co-Evolutionary Algorithm (CCEA), which aims to tackle a high-dimensional problem by decomposing it into two lower-dimensional search subproblems.
Moreover, we take a \textit{probabilistic} view of safe and unsafe regions and define a novel fitness function to measure the distance from the probabilistic hazard boundary and thus drive the search effectively.
We evaluate the effectiveness and efficiency of \MLCSHE on a complex Autonomous Vehicle (AV) case study.
Our evaluation results show that \MLCSHE is significantly more effective and efficient compared to a standard genetic algorithm and random search.

\end{abstract}

\begin{IEEEkeywords}
ML-enabled Autonomous System, Hazard Boundary, System Safety Monitoring, Cooperative Co-Evolutionary Search.
\end{IEEEkeywords}}

\maketitle

\IEEEdisplaynontitleabstractindextext

\IEEEpeerreviewmaketitle

\IEEEraisesectionheading{\section{Introduction}\label{sec:intro}}

\IEEEPARstart{A}{utonomous} systems are increasingly empowered by being embedded with ML components (MLCs) for various tasks, such as perception, localization, prediction, planning and control.
These components are inherently different from conventional software components and pose new challenges and safety risks that are not manageable by traditional software engineering practices. 
The main reason for this difference is that these components' logic is not captured by source code or specifications but their behavior is rather determined by training.
ML-enabled autonomous systems (MLASs) have already led to fatalities in the case of Autonomous Vehicles (AVs)~\cite{banerjee2018AVData}. 
This cannot be allowed to continue, especially when human life or very expensive equipment are involved.

Recent efforts have focused on making ML components more reliable, robust and accurate through novel testing methods~\cite{huang2020survey,zhang2020machine}.
However, even a system with reliable components can still lead to accidents~\cite{Leveson2012Engineering}. 
For example, some accidents are caused as a result of unsafe component interactions~\cite{albee2000report,Leveson2012Engineering}. 
Thus, the impact of ML components on safety can only be studied in the context of the system they are integrated into and in a specific operational context~\cite{black2009system, Leveson2012Engineering}.

The inherent specificity of ML components favors the use of safety monitors (also known as Run Time Assurance or RTA mechanisms)~\cite{skoog2020leveraging}. 
Safety monitors, at run time, check the inputs and outputs of a component that cannot be fully trusted, e.g., an ML component, and will block its outputs from being propagated to the rest of the system if they are potentially hazardous. 
In such cases, systems usually fall back on a trustworthy but less efficient component~\cite{asaadi2020assured}, or take any other pre-designed fallback mechanisms, such as stopping the AV on the shoulder of the road.
To do this, safety monitors have to observe the current state of the system and compare it with its Operational Design Domain (ODD)~\cite{J3016_202104}, to determine its deviation from ODD bounds since it might lead to hazards.
For instance, it is hazardous to rely on the self-driving feature of an AV on a rainy night if its ODD is characterized by normal dry operations during daylight. 
Additionally, safety monitors have to know the context of the system to determine whether the component might contribute to a hazard. 
For example, misclassification of an AV's object detection component might not lead to any hazards under a certain system context (henceforth called a \emph{scenario}), e.g., when an AV misidentifies an animal crossing the road as a pedestrian and stops.
Thus, identifying the combinations of system contexts (i.e., scenarios) and ML component's behaviors (i.e., inputs and outputs) that will transition the system to a hazard state is an essential step in developing safety monitors to be able to ensure the safety of the ML-enabled system.

However, there are several challenges involved with identifying the hazard boundary.
First, the problem space of scenarios and ML component behaviors is very large and high-dimensional and is thus a challenge for more conventional search metaheuristics such as Genetic Algorithms (GA). 
Second, the violation of a given safety requirement can only be determined if the system is executed within its operational environment, which involves computationally intensive simulations. The high computational cost, in addition to the large problem space, renders the problem even more challenging.
Last but not least, while safety can only be evaluated by executing the system within an environment, there are many environmental parameters that cannot be controlled even via a high-fidelity simulator; for example, the trajectory of pedestrians in CARLA~\cite{Dosovitskiy2017Carla}, a well-known AV simulator, is random.
Furthermore, two similar MLAS inputs may generate largely different outputs due to the non-linear behavior of ML models, such as Deep Neural Networks (DNNs).
Therefore, we \emph{cannot} assume that all combinations of scenarios and ML component behaviors within a region of the problem space have a uniform safety outcome, i.e., the region is deterministically safe or unsafe. Consequently, it is difficult to define hard boundaries between safe and unsafe regions. 

To address the aforementioned challenges, we propose \MLCSHE (ML Component Safety Hazard Envelope), a novel Cooperative Co-Evolutionary Algorithm (CCEA)-based approach that efficiently searches the problem space by decomposing it into two sub-spaces (one for scenarios and one for ML component behaviors) and parallelizing the search of sub-spaces while taking the joint contribution of both scenarios and ML component behaviors to the autonomous system safety into account. 
Moreover, instead of naively assuming that the hazard boundary is a clear line that exists between the safe and unsafe regions, we take a probabilistic view of the problem domain, i.e., at any point within the scenario and ML component behavior space, there is a probability of being safe. 
Based on this probabilistic lens, we present a novel fitness function that effectively guides the search towards the ``probabilistic'' hazard boundary based on the probability of finding safe scenario-behavior pairs within a given region.

\textit{Contributions.}
The contributions of this work are summarized as follows:
\begin{itemize}
    \item \MLCSHE, a dedicated and tailored cooperative coevolutionary search approach to approximate the hazard boundary of an ML component, in a probabilistic way, taking into account the combination of scenarios and MLC behaviors.
    \item An application of \MLCSHE to a complex Autonomous Vehicle (AV) case study involving an industry-strength simulator and an Autonomous Driving System with deep learning components.
    Our implementation of \MLCSHE as well as other case study artefacts are provided in our replication package (see Section~\ref{sec:data-availability}).
    \item An empirical evaluation of the effectiveness and efficiency of \MLCSHE through large scale experiments using CARLA, a high-fidelity open-source driving simulator, and Pylot, a high-performance open-source AV composed of multiple components.
    \item A comparison of \MLCSHE against baseline methods namely random search (RS) and vanilla genetic algorithm (GA).
\end{itemize}

\textit{Key Findings.}
The key findings of our empirical evaluation are summarized as follows:
\begin{itemize}
\item For reasonable boundary closeness thresholds given a search budget, \app is significantly more effective than RS and GA at detecting distinct boundary regions. This implies that a cooperative co-evolutionary algorithm makes the search for distinct boundary regions more effective than GA and RS.
\item For reasonable boundary closeness thresholds given a search budget, \app finds significantly more diverse regions that overlap with the hazard boundary at a faster rate than GA and RS. 
\end{itemize}

\textit{Paper Structure.}
The rest of the paper is structured as follows. 
Section~\ref{sec:background} provides background materials on CCEA. 
Section~\ref{sec:problem} defines the problem of MLC hazard boundary identification and details its challenges. 
Section~\ref{sec:relatedWork} discusses related work. 
Section~\ref{sec:coev} presents \MLCSHE in detail. Section~\ref{sec:evaluation} provides the empirical evaluation of \MLCSHE and discusses the results. 
Section~\ref{sec:conclusion} concludes and suggests future directions for research and improvement.

\section{Background}\label{sec:background}

In this section, an overview of Evolutionary Algorithms (EAs) is provided. Then we focus on a specific family of EAs, Cooperative Co-Evolutionary Algorithms (CCEAs), which happens to be particularly useful in our context.
Finally, the key decision points involved in designing a CCEA, namely collaborator selection and individual fitness assessment are discussed.

EAs are a family of algorithms designed based on the principles of evolutionary computation~\cite{Luke2013Metaheuristics}. EAs are inspired by the concepts related to biological evolution and have been applied to various optimization problems for which standard mathematical optimization is not applicable~\cite{Luke2013Metaheuristics}. EAs use concepts such as \textit{individual}, \textit{population}, \textit{fitness}, \textit{selection} and \textit{mutation} to formalize an optimization problem. Individuals usually represent solutions to the targeted problem and are members of a population whose fitness is evaluated (usually by a \textit{fitness function}). Desirable individuals, i.e., those with the highest fitness values, are more likely to be selected to act as the parents of the next generation. Using methods such as \textit{crossover} (replacing some parts of an individual with another one) and \textit{mutation} (adding randomness to some parts of an individual), individuals of the next generation population, i.e., next iteration of the search, are created.

For many problems, the search space is high-dimensional such that a conventional EA would not be able to solve it within a reasonable timeframe~\cite{ma2019survey}. To address this, Cooperative Coevolutionary Algorithms (CCEAs), originally proposed by \citet{CCEA-origin} in 1994, decompose the original problem into lower-dimensional subproblems, each of which can be solved in a separately evolving population as in conventional EAs described previously. Since individuals from each subproblem population must join together to form a \textit{complete solution} to the original problem, the fitness of an individual can only be evaluated based on the \textit{joint fitness} of the complete solution created by joining the individual with representative individuals, called \textit{collaborators}, from other populations. By carefully selecting collaborators and assessing individuals' fitness, CCEAs are known to be effective at solving even \textit{non-separable} problems~\cite{YANG20082985,panait2006archive} where the fitness of an individual of a subproblem population depends on the fitness of individuals of other populations. Furthermore, the decomposition of the original problem naturally allows parallelism to increase  search performance~\cite{ma2019survey}.

\begin{figure}[t]
\centering\includegraphics[width=\columnwidth]{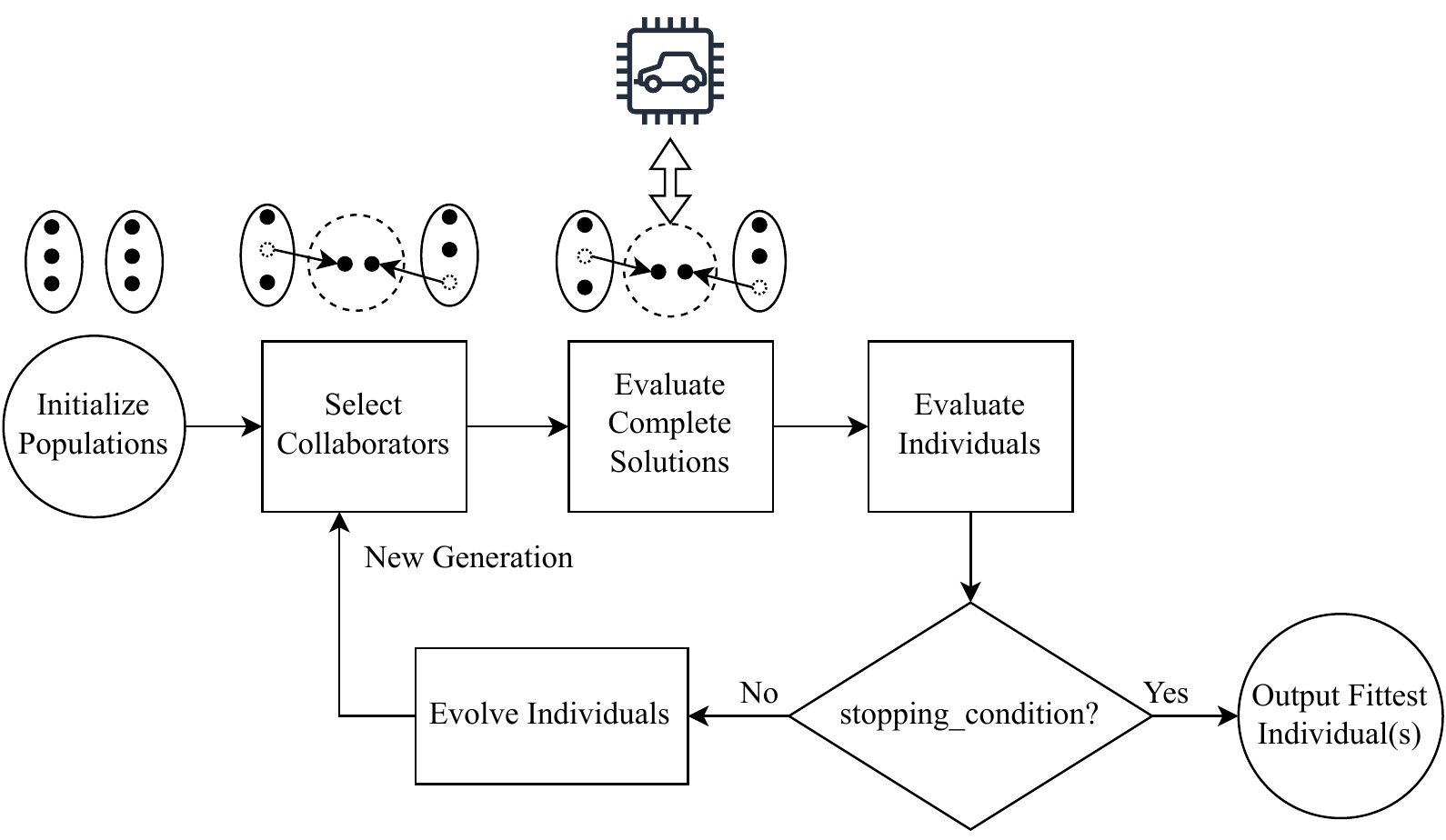}
\caption{An abstract coevolutionary algorithm (CCEA).}
\label{fig:CCEA}
\end{figure}

Figure~\ref{fig:CCEA} depicts the process of an abstract CCEA. Each population is initialized, either with randomly selected or guessed values (usually provided by domain experts). Individuals of each population collaborate with individuals of the other population(s) to form complete solutions. Then, these complete solutions are evaluated via joint fitness assessment functions. The joint assessments are then aggregated to provide evaluations of individual fitness values. If the \textit{stopping\_condition} is reached (true), then the fittest individuals are returned. Otherwise, the individuals go through breeding (selection, crossover and mutation) to create the next generation of the populations and go through evaluations again. 

Designing a CCEA includes two important decisions in the following aspects:
\textit{collaborator selection} and \textit{individual fitness assessment}.

\textit{Collaborator Selection.}
One of the most important factors affecting the performance of a CCEA is its \textit{collaborator selection} strategy.
To assess the fitness of the individual, the algorithm has to form one or multiple complete solutions with different collaborators. However, ideally, to get closer to the global optimum, all individuals of all other populations should be used as collaborators~\cite{panait2010theoretical}, which is usually infeasible due to resource constraints. Therefore, a strategy to efficiently select collaborators is required. Various strategies have been proposed in the literature, such as \textit{single best}, \textit{tournament-based}, and \textit{random}~\cite{ma2019survey}.
These strategies affect the algorithm via controlling the selection pressure and the pool size of the collaborators.

Some studies have proposed \textit{archive-based} collaborator selection to effectively reduce the number of collaborators to join in individual fitness assessments while maintaining the amount of information contained in the populations~\cite{ma2019survey}. The idea is to carefully select a \emph{population archive} which is a subset of a population to be used as collaborators. 
For example, Panait et al.~\cite{panait2006archive}, have proposed \emph{iCCEA}, which aims to minimize the size of the population archives by considering only the collaborators that are \textit{informative} and \textit{distinct}. A collaborator in an archive is informative if adding it to the archive changes the fitness ranking of the population's individuals. If there are multiple collaborators that can change the ranking of the same individuals, the collaborator that changes the ranking the most will be kept in the archive. A collaborator in an archive is distinct if its (Euclidean) distance from other collaborators in the archive is higher than a pre-defined threshold. As a result, a population archive keeps only a minimum number of collaborators while attempting not to lose information in terms of collaborations between subproblem solutions. 
However, though the population archive selected by iCCEA is minimal in size, the algorithm proposed by the authors to update the population archive in each generation has a high time complexity ($O(n^3)$ where $n$ in the number of individuals in the archive) and this severely impacts the performance of the algorithm. 
Thus, simpler population archive selection methods, e.g., \emph{elitist}, \emph{random} and \emph{best+random}, that are much faster, are also widely used in practice.

\textit{Individual Fitness Assessment.}
The \textit{collaborator selection} strategy of a CCEA affects its \textit{individual fitness assessment} strategies as well. The only \textit{objective} fitness assessment that can be done on the individuals is based on their joint fitness assessments with collaborators. Thus, all algorithms perform some form of aggregation on joint fitness assessments related to an individual to determine its fitness value. \textit{Best}, \textit{worst} or \textit{average} joint fitness values are usually used for individual fitness assessments.

\section{Problem and Challenges}\label{sec:problem}

In this section, we provide a precise problem definition regarding the identification of the boundaries of hazard envelopes, focusing on the behavior of a Machine Learning Component (MLC). While we use an AV as an example, it can be easily generalised to any ML-based Autonomous System (MLAS).

\subsection{Problem Definition}\label{sec:prob-def}

\begin{figure}
\centering\includegraphics[width=\linewidth]{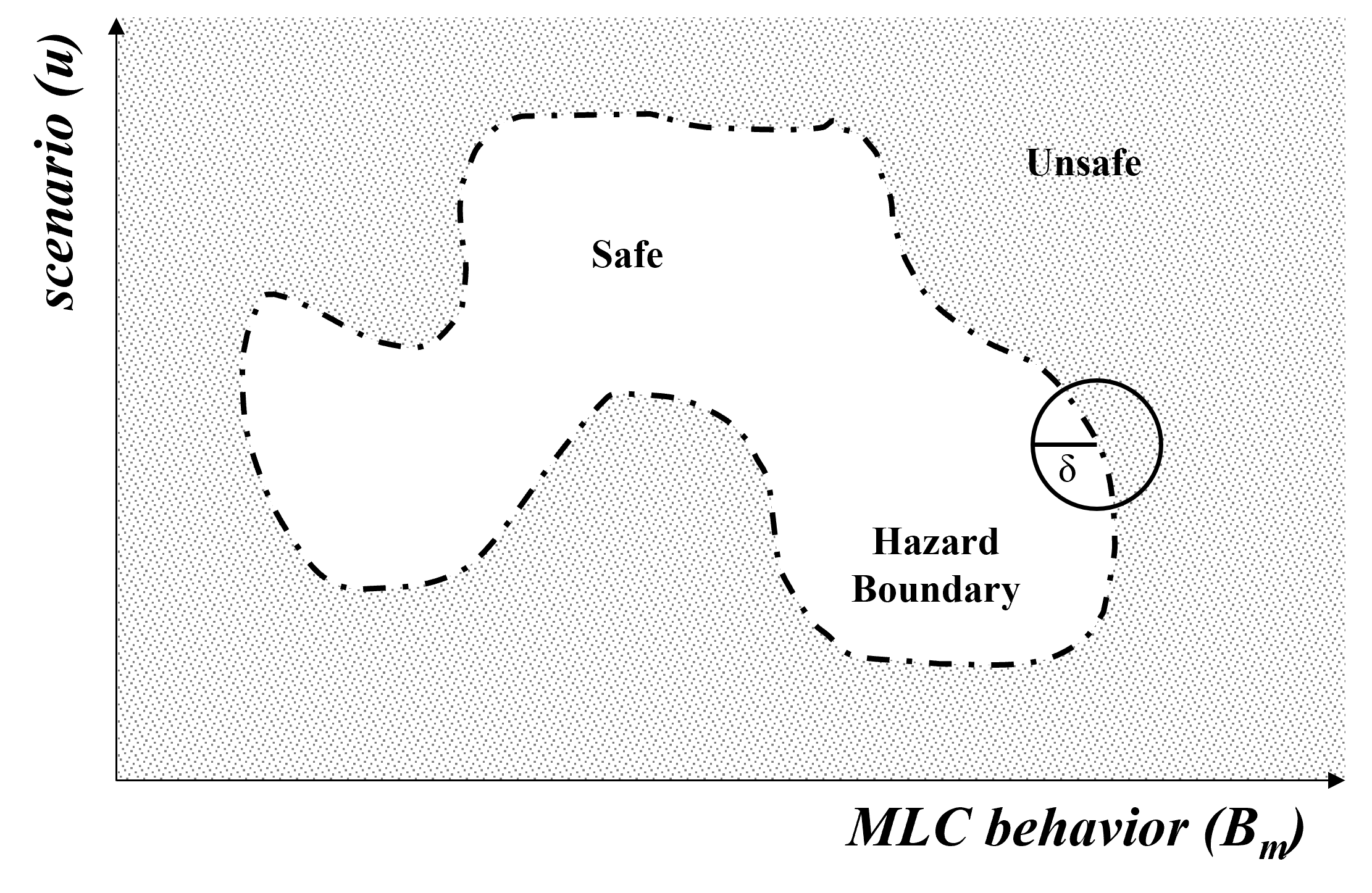}
\caption{An illustration of safe and unsafe regions and the corresponding hazard boundary.}
\label{fig:ProblemSpace}
\end{figure}

Consider an AV as an ML-enabled Autonomous System (MLAS) including an ML component (MLC), namely an image-based object detection component using DNNs. 
The AV continuously observes its surrounding environments---such as roads, traffic signs, buildings, and other moving vehicles via sensors (e.g., camera) and generates driving commands (e.g., steer left and decrease speed) to best satisfy given functional and safety requirements (e.g., reach a given destination point without colliding with other vehicles). During testing of the AV, the environment is often simulated by a high-fidelity driving simulator due to the high cost and risk of real-world testing.
Inside of the AV, whenever new sensor data (e.g., an image taken from the camera) is collected, it passes through the object detection component to identify the positions of surrounding objects, if any, from the (fused) sensor data (e.g., in the form of bounding boxes in the given image), which will then be used to determine proper driving commands.
Under a certain driving scenario, the AV might violate requirements such as ``\textit{the AV shall keep a minimum distance of \SI{1.5}{\meter} from any vehicle in front.}'' In such cases, the MLC could internally contribute to the violation.
Therefore, identifying the boundaries of the hazard envelopes of the AV in terms of the combination of driving scenarios and MLC behaviors is important.
Figure~\ref{fig:ProblemSpace} provides a simplified illustration of an ML component's hazard envelope, defined in terms of scenarios and ML component behaviors, where a \textit{safe} region leading to no violations is surrounded by an \textit{unsafe} region leading to violations.
The goal is to identify, as precisely and completely as possible, the boundaries between safe and unsafe regions, illustrated by the dashed line in Figure~\ref{fig:ProblemSpace}.

More specifically, let $s$ be the AV including the MLC $m$ for image-based object detection, operating in a simulated driving environment.
For a given scenario $u$, which consists of all the static and dynamic entities of the environment such as road shape, weather, and other vehicles, 
the simulation result for $s$ and $m$, denoted by $\Pi_{u, s, m}$, is a sequence $\langle e_1, e_2, \dots, e_T \rangle$ where $T$ is the duration of the execution and $e_t$ for $t=1,\dots,T$ is the snapshot (state) of the environment at time step $t$.
For each time $t\in \{1,\dots,T\}$, $s$ takes an image $in_{s,t}$ taken from the camera by observing $e_t$ and generates a pre-processed (e.g., gray-scaled) image $in_{m,t}$ for $m$.
Then, $m$ produces the object information $out_{m,t}$ (in the form of bounding boxes) by processing $in_{m,t}$ and $s$ produces driving commands $out_{s,t}$ by processing $out_{m,t}$\footnote{Note that there can be other components in $s$ that interact with $m$. For example, $out_{m,t}$ can be one of the (possibly many) factors that determine $out_{s,t}$. We only assume that $m$ is one of the (possibly many) components required for $s$ to operate; specifically, $in_{s,t}$ can affect $in_{m,t}$ and $out_{m,t}$ can affect $out_{s,t}$.}. 
The environment snapshot $e_{t+1}$ for the next time step $t+1$ is updated based on $out_{s,t}$, and the whole process repeats until $t$ reaches $T$.
The behavior of $m$, denoted by $B_m$, is defined as the sequence of input/output pairs such that, for an input/output pair $(in_m, out_m) \in B_m$, $out_m$ is the output produced by $m$ by processing $in_m$.
For a safety requirement $r$ (e.g., do not collide with other vehicles), we can measure the degree of the safety violation (e.g., the distance to the other vehicles) of $s$ for $u$ and $B_m$ in terms of $r$, denoted by $f(r, \Pi_{u, s, m})$, by analyzing $\Pi_{u, s, m} = \langle e_1, e_2, \dots, e_T \rangle$ against $r$.
If $f(r, \Pi_{u, s, m}) > \epsilon$ for a small threshold $\epsilon$ predefined for $r$, we say that $r$ for $s$ is violated by (the combination of) $u$ and $B_m$.
This means that we can decide the violation of $r$ (with $\epsilon$) given $u$ and $B_m$.

Given the above context, let $(u, B_m)$ be a point in a space referred to as the \textit{input space}, that is defined by (the combination of) possible scenarios and MLC behaviors.
For each point $(u, B_m)$ in the input space, we can decide its output (i.e., unsafe or safe) by checking whether it leads to the violation of $r$ or not.
The identification of the boundaries of hazard envelopes attempts to find as many $(u, B_m)$ points as possible that are close to the boundaries between safe and unsafe regions in the space.
Notice that we intentionally left the precise definition of safe and unsafe regions unclear since it is one of the challenges we address next.

\subsection{Challenges}\label{sec:prob-challenges}

The problem of hazard boundary identification for an MLC in the MLAS under analysis, entails multiple major challenges.

As discussed in Section~\ref{sec:prob-def}, both a scenario $u$ and an MLC behavior $B_m$ collaboratively determine the violation or satisfaction of a safety requirement $r$.
As a result, there are too many possible scenarios and MLC behaviors for the input space to be exhaustively explored without resorting to limiting assumptions that can bias the results~\cite{Norden2019sampling}.
One might argue that unsafe regions of the input space could be analytically identified using methods based on expert knowledge, such as FTA~\cite{standard2006FTA} and HAZOP~\cite{standard2001hazop}, to provide clear insights into how hazards can occur.
However, such methods are not sufficient to address all possible ways hazards can arise due to
complex interactions between MLAS components and the opacity of ML components. 

Second, the satisfaction or violation of $r$ can only be determined if the system is operated within its surrounding environment. 
During testing, in addition to the first challenge above, this requires running a high-fidelity simulator which is generally very resource-intensive. 
The high cost of simulation highlights the need for an efficient and effective method to search as much of the input space as possible while focusing on the regions close to the boundary. 

Lastly, recall it is unrealistic to consider a region 100\% safe or unsafe.
This is explained by two main reasons. 
First, simulators do not often enable full control of all relevant parameters in the environment, thus randomly configuring some of them. For example, the movement of pedestrians is random in CARLA~\cite{CarlaAPI}, a high-fidelity simulator. 
Second, two inputs that are close in the input space may generate different MLC outputs that are handled differently by the rest of the system, e.g., due to the non-linear behavior of other DNNs using the MLC outputs as their input, resulting in different safety results (i.e., safe or unsafe). 
As a result, we cannot assume a uniform and consistent safety outcome for a region, making it difficult to define hard boundaries between safe and unsafe regions. 
Rather, hazard envelope boundaries (i.e., the dashed line in Figure~\ref{fig:ProblemSpace}) should be probabilistic as they encompass regions with a given probability threshold of violating a selected requirement. 

To address the above challenges, we propose a novel method using Cooperative Co-Evolutionary Algorithm (CCEA) that efficiently address our objectives as an optimization problem, within a large input space, by decomposing such problem into lower-dimensional subproblems. 
Further, to recast our problem into a coevolutionary search problem, we define a special fitness function that can assess how far a candidate solution (i.e., a combination of $u$ and $B_m$) is from the boundary of a ``probabilistic'' unsafe region. 
See Section~\ref{sec:coev} for details of our method.

\section{Related Work}
\label{sec:relatedWork}

This section discusses existing studies related to the problem of hazard envelope boundary identification. 
Depending on the methods used, we found three categories: search-based methods, sampling-based methods, and formal methods. 

\subsection{Search-based Methods}
Search-based methods employ metaheuristics (search algorithms) and convert the boundary identification problem into a search problem guided by a fitness function that evaluates how close a system input (e.g., test scenario) is from the boundary. Fitness assessment for individual system inputs often involves simulation executions to check whether safety requirements are violated. 

\begin{figure}
\centering
\includegraphics[width=\linewidth]{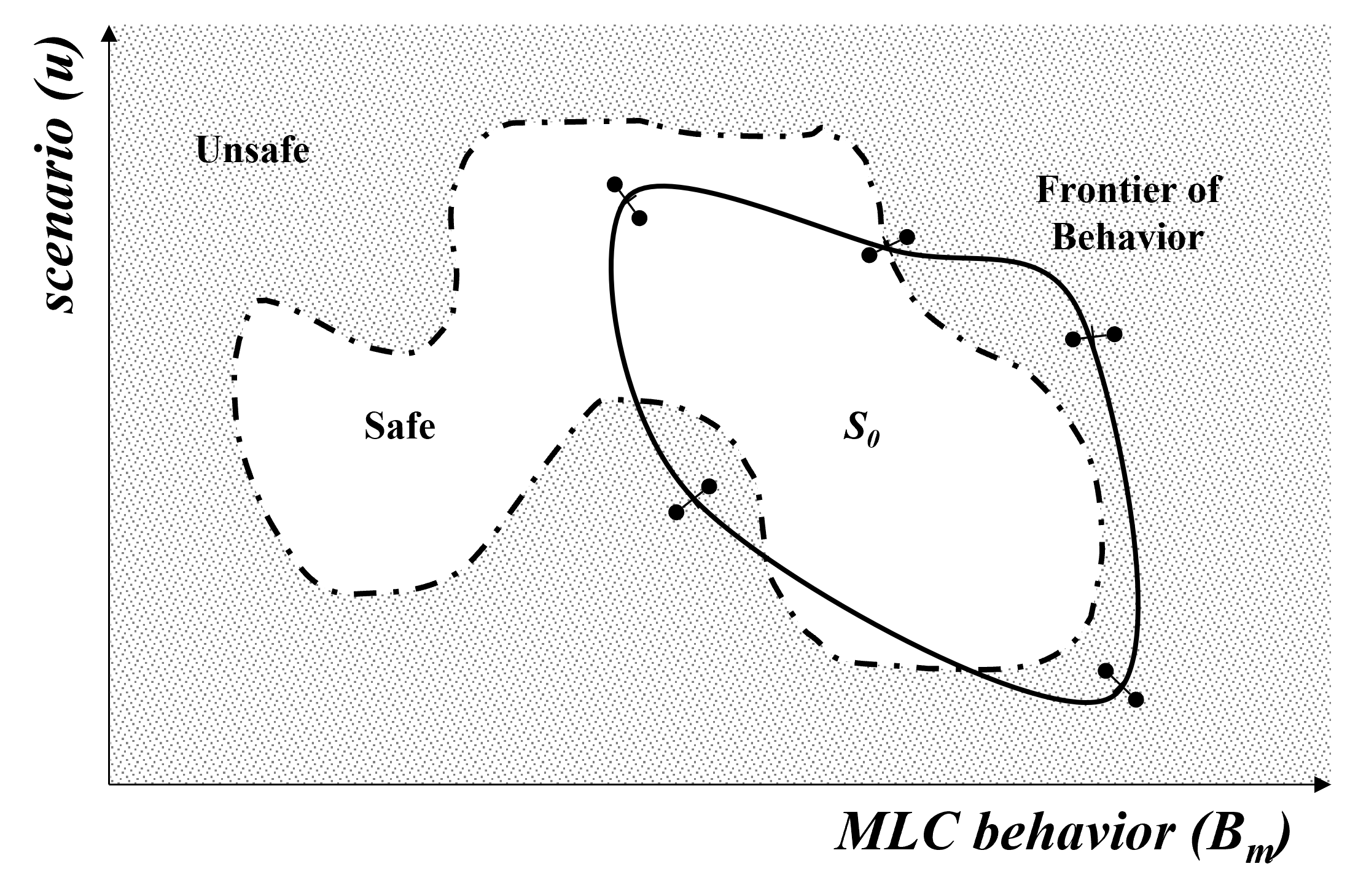}
\caption{A possible application of \emph{DeepJanus} to the systemic hazard boundary detection problem. Connected dots are a safe and unsafe pair.
}
\label{fig:DeepJanus}
\end{figure}

Although there are many search-based methods for testing MLCs~\cite{zhang2020machine, riccio2020testing}, the problem of boundary identification has received very little attention. 
Only recently, \citet{riccio2020deepjanus} proposed DeepJanus, the first search-based method to identify the \textit{frontier of behavior} (frontier) of MLCs, i.e., a set of similar input pairs that trigger different behaviors (e.g., safe and unsafe) of the system. 
The discovered frontier can allow developers to approximate a safe operating envelope for the MLC (by interpolating the pairs).
Also, the overlap of the estimated safe operating envelope with the \textit{validity domain}
of the MLC, which is the domain where the MLC is expected to behave according to its requirement(s)~\cite{riccio2020deepjanus}, can facilitate the evaluation of the MLC's quality.
Therefore, for example, DeepJanus can be useful in distinguishing between the performance of two MLCs that perform the same task.
However, it cannot solve the issue of identifying the hazard boundary, as the impact of MLCs on safety can only be assessed when evaluating the entire system in a given environmental context.
Furthermore, as illustrated in Figure~\ref{fig:DeepJanus}, the interpolated frontier of behavior and the hazard boundary of an MLC are not necessarily the same. 
More precisely, a member of the frontier (i.e., a pair of safe and unsafe inputs) does not necessarily lie in proximity to the hazard boundary since the violations can occur in probabilistic safe regions (i.e., regions where the proportion of safe inputs is above a certain threshold) as argued in Section~\ref{sec:prob-challenges}.
Therefore, we need a novel method to identify the hazard boundary of an MLC within a system, considering the probabilistic nature of (un)safe inputs. 

\subsection{Sampling-based Methods}
Unlike search-based methods, which are guided by fitness functions, sampling-based methods use repeated random samplings (e.g., Monte Carlo methods) or statistical metrics to identify certain system inputs that lead to safety violations.
For example, Meltz and Guterman~\cite{Meltz2019} proposed SmARTest, which uses Monte Carlo methods to identify a \textit{scenario domain} (i.e., set of system inputs) that lead to safety requirement violations determined by measuring \textit{Performance Assessment Functions} (PAFs) defined based on the requirements' Key Performance Indicators (KPIs).
\citet{Sinha2020} proposed Neural Bridge Sampling (NBS), a method to measure the probability of rare events, such as accidents, using Monte Carlo methods. NBS decomposes the probability of a rare event into chained conditional probabilities, which are tractable to compute using standard Monte Carlo methods. This provides a better estimate than the naive Monte Carlo or Adaptive Multi-Splitting (AMS) methods.

Sampling-based methods can efficiently identify safe and unsafe inputs from the system's input space. However, they only consider system inputs (i.e., scenarios) and not the effect of different MLC behaviors for the same scenario. As discussed in Section~\ref{sec:prob-challenges}, both the scenario and the MLC behavior must be taken into account to determine the conditions when MLC behavior leads to safety violations.

\subsection{Formal Methods}
Formal methods rely on formal representations of the input space, the system (including the MLC), and the output space. Examples of such representations include hybrid system or dynamical system formalisms~\cite{antsaklis1992hybrid}. Tools like SMT solvers~\cite{katz2017reluplex} and Mixed Integer Linear Programming (MILP)~\cite{lomuscio2017milp} can be used to analyze whether the system containing the MLC can reach an unsafe region given its input space~\cite{huang2020survey}. This is known as \emph{reachability analysis}.
Reachability analysis is used to identify the MLC's \emph{barrier certificate}, which is an invariant function that constrains the state space of the system and ensures the satisfaction of a safety property~\cite{barrierCertificates} while considering the closed-loop behavior of the system. Barrier certificates can be seen as an over-approximation of the hazard boundary of the MLC.

\citet{ivanov2019verisig} proposed Verisig, which can be applied to Cyber-Physical Systems (CPSs) with DNN-based feed-forward controllers with ReLU activation functions. Verisig transforms the ReLU DNN into a hybrid system representation and combines it with the rest of the system. This recasts the problem as a hybrid system verification problem. Given a set of system inputs, the outputs can be approximated using Flow$^{*}$~\cite{Chen2013Flow}, a nonlinear system reachability analyzer.
\citet{Tuncali2018} proposed another method to identify barrier certificates of DNN-based, feed-forward controllers, which is not limited to architectures with ReLU activation functions. This method first identifies candidate barrier certificates using simulations, then evaluates their suitability using the dReal~\cite{gao2013dreal}, an SMT solver for nonlinear formulas in real numbers.
\citet{Tran2019NNV} proposed NNV, a method to perform closed-loop reachability analysis of control systems with Deep Reinforcement Learning (DRL) controllers. These controllers have a feed-forward architecture with ReLU/Saturation activation functions. NNV calculates a low-error over-approximation of the output region, which are reached by the system given its inputs.

Although the aforementioned methods provide guarantees for the hazard boundary and cover all possible trajectories of the system, they suffer from practicality and scalability issues.
For example, over-approximation of the hazard boundary might incorrectly reduce (or even remove in the worst case) the safe operating envelope of the system by incorrectly considering some safe behaviors unsafe, thus limiting the practicality of the methods~\cite{Tran2019reachability}.
Furthermore, reachability analysis can only be applied to feed-forward controllers with specific activation functions. Thus, it cannot be used for practical MLCs that perform perception, obstacle tracking, or prediction tasks with different DNN architectures (e.g., recurrent neural networks). Also, reachability analysis has not yet been applied to closed-loop, industrial Cyber-Physical Systems (CPS) with feedback DNN controllers~\cite{ivanov2019verisig, Tran2019NNV, Tuncali2018}. In such a context, scalability is very likely to become an acute problem. 

\subsection{Remark on Differences in Objectives}
A common goal underlying all the above-mentioned methods is to identify the hazard boundary of a \textit{given} MLC embedded within its containing system (MLAS). 
It could be useful when the MLC under test is fixed, but as soon as the MLC changes (e.g., via retraining), the previously identified hazard boundary would be invalid, and the whole safety verification exercise would have to be repeated.
On the other hand, in our research, we aim to identify the combinations of conditions and MLC behaviors, \textit{without} referring to a specific MLC implementation, that could potentially lead to hazards. Once characterized, such situations could then, independently of a specific MLC implementation, be used to monitor the operation of the system and MLC and warn the user in case it is operating near to the hazard boundary.

In the following section, we propose a novel method that addresses the challenges discussed in Section~\ref{sec:prob-challenges}, and is applicable to various types of MLCs, such as perception, planning, and control, without making any assumptions about their architecture.

\section{Our Approach}\label{sec:coev}

In this section, we provide a solution to the problem described in Section~\ref{sec:problem}, i.e., the hazard boundary identification of an MLC in the MLAS under analysis. 
Our key idea is to recast the problem as a cooperative co-evolutionary search problem where scenarios and MLC behaviors co-evolve as two separate populations but contribute together to find complete solutions (i.e., the combinations of scenarios and MLC behaviors) close to the boundary. 
Then, we use CCEAs, the algorithms that are well known to be effective at solving search problems such as the one described in Section~\ref{sec:background}.

In the following subsections, we first describe how scenarios and MLC behaviors can be represented as two separate populations in a search problem (Section~\ref{sec:reps}). 
We then define a novel fitness function of the search problem to assess how close a complete solution is from the boundary (Section~\ref{sec:fitnessFunc}).
Finally, we present our novel method based on CCEAs using the representation and the fitness function (Section~\ref{sec:algorithm}).

\subsection{Representations}
\label{sec:reps}

We consider two populations, one for scenarios and another one for MLC behaviors.
This is to consider \textit{all possible} MLC behaviors that can lead to the boundary regions when combined with certain scenarios. Note that \app does not aim to test a particular MLC in the system under analysis, but rather to monitor the behavior of current and future implementations of the MLC using the resulting boundary information. Therefore, it is important to manipulate MLC behaviors and scenarios to find all boundary regions.
However, the individuals of the MLC population, subjected to evolutionary operators, are only represented as \textit{MLC-outputs} ($out_m$).
This is due to the initial \textit{MLC-input} ($in_m$) being (indirectly) determined by the scenario, whereas the next MLC-inputs are affected by previous MLC-outputs.
Therefore, $in_m$ is recorded in an \textit{archive} of complete solutions (i.e., $A_c$ in Algorithm~\ref{alg:MLCSHE}; see Section~\ref{sec:algorithm} for details) but not included in the representation of the behavior of the MLC that can be manipulated by the search.
Recording the $in_m$ and $out_m$ sequences, along with their corresponding scenarios ($u$), is indeed crucial as it records unsafe behaviors of an MLC (its input and output sequences) given a set of environmental conditions (its \textit{scenarios}). This information enables the design of safety monitors that will prevent the MLC from contributing to a systemic hazard via leveraging the recorded information.

\subsubsection{MLC behaviors}\label{sec:rep:mlc-output}
One of the two populations considered for the search is the set of MLC behaviors. The behavior of an MLC can be expressed as a sequence of input and output tuples. However, as discussed above, the inputs of an MLC are indirectly controlled by the environmental input to the system (i.e., scenario parameters) and the components of the system that process that input before it is passed on to the MLC. Thus, the parameters that we can directly manipulate during the search are the outputs of the MLC. We represent an individual in the population of the MLC behaviors as a sequence of MLC outputs where the $t$-th element of the sequence denotes an MLC output at time step $t$.

The output of an MLC depends on the task performed by the MLC. 
For instance, in the case of a steering angle estimator, the output is a single real value. 
Whereas, in the case of an object classifier, the output is a vector of probabilities (real values between 0 and 1), where each element corresponds to a label. 
Finally, similar to our running example, in the case of obstacle detection, the outputs in an ML component (MLC) are detected obstacles, i.e., their bounding box\footnote{A bounding box specifies the area on the image processed by the obstacle detector that contains the detected obstacle. It can be expressed as ($x_{min}$, $x_{max}$, $y_{min}$, $y_{max}$) corresponding to a specific 2D box.}, their label (such as pedestrian, vehicle, lamp post, etc.), and their timestamp. 
Therefore, an $\mathit{mlco}$ (MLC Output) for a simulation duration $T$ can be defined as a sequence of the trajectories of detected obstacles during $T$ in the case of obstacle detection.

Specifically, given the maximum number of detectable objects $n$ and the simulation duration $T$, an $\mathit{mlco}$ can be defined as a sequence of $n$ trajectories $\langle \mathit{trj}_1, \dots, \mathit{trj}_n \rangle$ where $\mathit{trj}_i$ represents the trajectory of the $i$-th object (in terms of the bounding boxes) for $T$.
By allowing the search algorithm to manipulate individual trajectories, an arbitrary $\mathit{mlco}$ can be generated for obstacle detection.

However, allowing the search algorithm to generate all the bounding boxes for individual time steps will likely yield an unrealistic trajectory randomly moving around without a consistent direction, which we observed during our initial trials.
Therefore, it is better to allow the search algorithm to generate only the \textit{start} and \textit{end} bounding boxes, and then generate the remaining bounding boxes for intermediate time steps using linear interpolation between the start and end boxes.
Specifically, $\mathit{trj}_i$ can be defined as a triple $(\mathit{class}_i, \mathit{start}_i, \mathit{end}_i)$ 
where $\mathit{class}_i$ is the class of the $i$-th object (e.g., car, bicycle, pedestrian), 
$\mathit{start}_i$ is the position and the size of the bounding box of the $i$-th object at time step $t=t_{\mathit{start}}$, 
and $end_i$ is the position and the size of the bounding box of the $i$-th object at time step $t=t_{\mathit{end}}$.
For example, \textit{start} or \textit{end} can be defined as a quintuple ($t$, $x_{min}$, $x_{max}$, $y_{min}$, $y_{max}$), which are time and bounding box parameters for the beginning or the end of the trajectory, respectively. 
Then, for a given trajectory $\mathit{trj} = (\mathit{class}, \mathit{start}, \mathit{end})$, we can easily generate the positions and sizes of bounding boxes for intermediate time steps (i.e., $1 < t < T$) based on \textit{start} and \textit{end} (using linear interpolation) whenever needed for a simulation.

We want to note that \app does not aim to test a given MLC implementation in the system under analysis, but to monitor the behavior of any current or future implementation of the MLC using the resulting boundary information (see Sections~\ref{sec:intro} and \ref{sec:prob-def}).
For a given implementation, there may indeed be ``unfeasible-in-practice'' MLC behavior for a given scenario, which is, however, hard to determine beforehand. But then this is part of the problem space that will never be reached at run-time and is not an issue.

\subsubsection{Scenarios}
\label{sec:rep:scenario}
A scenario can be represented as a heterogeneous vector of real and integer values. For the case of an AV, a scenario consists of the vehicle itself, the weather, the road and other static (e.g., lamp posts and other obstacles) and dynamic objects (e.g., pedestrians and other cars)~\cite{haq2021can}. Each of them have many attributes of various types, namely float (e.g., speed) and enumerated types (e.g., line pattern) which can be encoded as integer values.

The size of a scenario individual is determined by the simulator. Furthermore, a finer-grained level of simulation control implies a larger scenario size as more parameters have to be manipulated by the search algorithm. For instance, one can manipulate all weather-related parameters separately (10 parameters in the case of CARLA~\cite{CarlaAPI}) or manipulate them using the weather preset parameter (1 parameter) which sets the value of all granular weather parameters according to high-level modalities, e.g., rainy sunset, clear noon.

\begin{figure*}[ht]
\centering\includegraphics[width=0.8\linewidth]{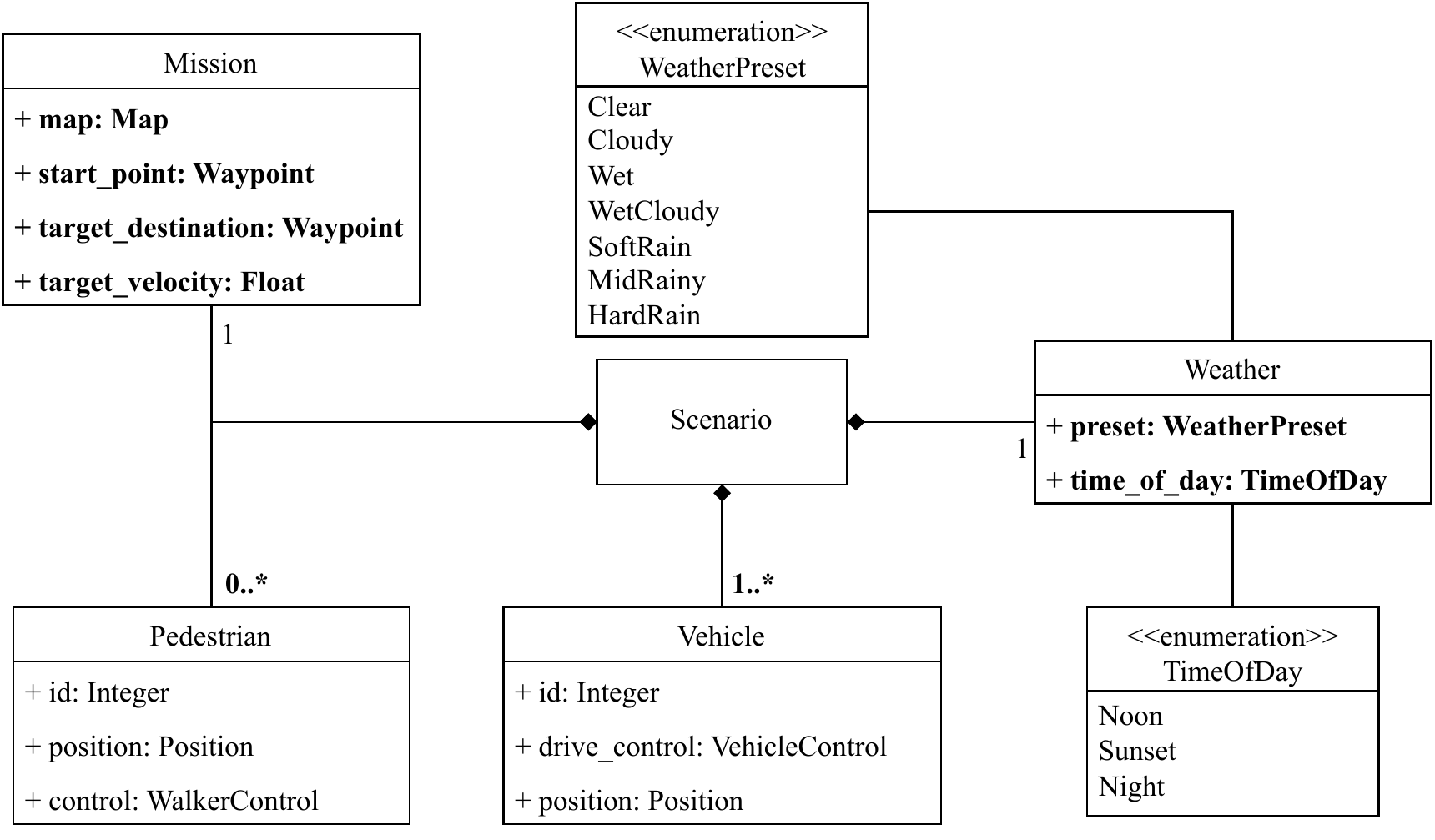}
\caption{The scenario domain model for the running example. The model is based on the concepts provided in the Carla World domain model~\cite{Dosovitskiy2017Carla, CarlaAPI}. The scenario parameters are shown on the figure in bold font, i.e., the weather \textsf{preset}, the attributes of \textsf{Mission} and the number of actors such as \textsf{Vehicle}s and \textsf{Pedestrian}s.}
\label{fig:ScenOnt}
\end{figure*}

Figure~\ref{fig:ScenOnt} is the scenario domain model for our running example. A \textsf{Scenario} consists of one or more \textsf{Vehicle}s (including the ego vehicle), zero or more \textsf{Pedestrian}s and, \textsf{Mission} and \textsf{Weather}. The attributes of the domain model that act as the parameters for a scenario representation are written in bold font in Figure~\ref{fig:ScenOnt}. Therefore, a scenario can be defined by the \textsf{time\_of\_day}, weather \textsf{preset}, \textsf{map} of the town, \textsf{start\_point} of the ego vehicle, its \textsf{target\_destination} and \textsf{target\_velocity}, the number of \textsf{Pedestrian}s, and the number and \textsf{position} of other \textsf{Vehicle}s with respect to the ego vehicle (e.g., in front, on the opposite lane).

\textit{Operational Design Domain (ODD).}
The Operational Design Domain or ODD defines an operational envelope of the AV, i.e., a set of bounds on the environmental parameters of the system. For instance, highway driving is an ODD for AVs which determines the type of the road, the average speed of the surrounding vehicles, and the (lack of) pedestrians in the vicinity~\cite{J3016_202104}. However, within an ODD, many scenarios can still be defined, e.g., weather, the number of cars, the length and shape of the road. Therefore, a search can be done within an ODD, which sets the values or the bounds of some parameters, such as the target speed of the ego vehicle. The parameter bounds or values set by the ODD will remain static for the duration of the search, e.g., a target speed of $90$kph in a highway driving ODD, or the angle of the sunlight during a daytime driving ODD.

\subsection{Fitness Function}\label{sec:fitnessFunc}

This section presents our proposed fitness function in detail. Our aim is to design a fitness function that can effectively guide the search towards the boundary of unsafe regions. 
However, as mentioned in Section~\ref{sec:problem}, we cannot assume that a region is either 100\% unsafe or safe. To address this, we first define the notion of safe and unsafe inputs, followed by \textit{probabilistic} unsafe regions.

\begin{definition}[Safe and Unsafe Inputs]
An input is \textit{unsafe} if and only if it leads to the violation of a given requirement. Otherwise, the input is \textit{safe}. 
\end{definition}

Recall that an input is a combination of a scenario and an MLC behavior in our context. 

\begin{definition}[Probabilistic Unsafe Region]
Let $X$ be a set of all possible inputs, representing the input space. 
Given a threshold probability $p_\mathit{th}$, a region $G\subseteq X$ is \textit{$p_\mathit{th}$-unsafe} when the proportion of unsafe inputs in $G$ is higher than $p_\mathit{th}$.
\end{definition}
For example, if we randomly draw an input from a 5\%-unsafe region, we have more than 5\% chance of leading to a safety violation. The value of $p_\mathit{th}$ should be determined by a domain expert within a specific application context.

Notice that the shape of a probabilistic unsafe region is unknown, as is its boundary. 
Nevertheless, we can \textit{approximate} how far an arbitrary input is from the boundary by sampling its neighborhood. 
Specifically, for an input $x\in X$, let $p_x$ be the proportion of unsafe inputs in the neighborhood of $x$.
If $p_x \le p_\mathit{th}$, it implies that $x$ is not likely to be located in a $p_\mathit{th}$-unsafe region.
Otherwise, if $p_x > p_\mathit{th}$, it implies that $x$ is likely in a $p_\mathit{th}$-unsafe region.
Therefore, if $p_x$ is close to $p_\mathit{th}$, it implies that $x$ is close to the boundary of a $p_\mathit{th}$-unsafe region. 
To leverage this idea, we define the notion of neighborhood as follows:
\begin{definition}[Neighborhood]\label{def:neighborhood}
For an input $x \in X$ and a non-negative real number $\delta \in \mathbb{R}^+$, a \textit{neighborhood} of $x$ with the radius of $\delta$, denoted by $N(x, \delta)$, is defined as follows:
\begin{equation}
N(x, \delta) = \{x'\in X |~\textit{dist}(x, x') \le \delta \}
\end{equation}
where $\textit{dist}(x, x')$ indicates the distance between $x$ and $x'$.
\end{definition}
Notice that various distance functions \textit{dist} can be adopted depending on the nature of complete solutions.
For example, if a complete solution can be represented as a heterogeneous vector composed of numerical, ordinal, and categorical values, \textit{heterogeneous distance metrics} \cite{wilson1997improved} are good candidates to measure the distance between two complete solutions.
In Figure~\ref{fig:ProblemSpace}, a neighborhood with a radius of $\delta$ is visualised as a circle between safe and unsafe regions.

Based on Definition~\ref{def:neighborhood}, let $p_{x, \delta}$ be the proportion of unsafe inputs in $N(x, \delta)$. 
Then, as discussed above, we can use the difference between $p_{x, \delta}$ and $p_\mathit{th}$ to approximate the distance between $x$ and the boundary of a $p_\mathit{th}$-unsafe region.
However, we cannot compute the exact value of $p_{x, \delta}$ since $N(x, \delta)$ has too many complete solutions to exhaustively evaluate.
Nevertheless, we can compute an \textit{estimate} of $p_{x, \delta}$, denoted by $\hat{p}_{x, \delta}$, and its confidence interval since the consecutive trials of checking whether an input $x' \in N(x, \delta)$ is safe or not are assumed to be independent and can be treated as Bernoulli Experiments.

Specifically, the probability distribution of $p_{x, \delta}$ can be modelled as a Binomial distribution, and we can compute $\hat{p}_{x, \delta}$ as follows:
\begin{equation}\label{eq:probability}
    \hat{p}_{x, \delta} = \frac{\mathit{unsafe}(N(x, \delta))}{\mathit{evaluated}(N(x, \delta))}
\end{equation}
where $\mathit{evaluated}(N(x, \delta))$ is the number of inputs evaluated (sampled) in $N(x, \delta)$ and $\mathit{unsafe}(N(x, \delta))$ is the number of unsafe inputs among those evaluated. 
Furthermore, using the Wilson Confidence Intervals~\cite{brown2001interval}, we can compute the confidence interval of $p_{x, \delta}$, denoted by $\mathit{CI}(p_{x, \delta})$, as follows:
\begin{equation}\label{eq:confInterval}
    \begin{split}
        & \mathit{CI}(p_{x, \delta}) = \frac{1}{1+\gamma} \big(\hat{p}_{x, \delta} + \frac{\gamma}{2}\big) \\
        & \pm \frac{z}{1+\gamma}\sqrt{\frac{\hat{p}_{x, \delta}(1 - \hat{p}_{x, \delta})}{{\mathit{evaluated}(N(x, \delta))}}+\frac{\gamma}{4\times{\mathit{evaluated}(N(x, \delta))}}}
    \end{split}
\end{equation}
where $\gamma = \frac{z^2}{{\mathit{evaluated}(N(x, \delta))}}$ and $z$ is determined by the standard normal distribution for a given confidence level (e.g., for a $95\%$ confidence level, $z=1.96$).

Based on $\mathit{CI}(p_{x, \delta})$, we can assess the maximum difference\footnote{We consider the maximum difference to be conservative.} between $p_{x, \delta}$ and $p_\mathit{th}$ as follows:
\begin{equation}
\mathit{diff}(p_{x, \delta}, p_\mathit{th}) = \max \big(|\mathit{UL}(p_{x, \delta}) - p_\mathit{th}|, |\mathit{LL}(p_{x, \delta}) - p_\mathit{th}|\big)
\end{equation}
where $\mathit{UL}(p_{x, \delta})$ and $\mathit{LL}(p_{x, \delta})$ are the upper and lower limits of $\mathit{CI}(p_{x, \delta})$, respectively. 
Using $\mathit{diff}(p_{x, \delta}, p_\mathit{th})$, we define our fitness function as follows.

\begin{definition}[Boundary-Seeking Fitness Function]\label{def:ff}
For an input $x$, a neighborhood radius $\delta$, and a threshold probability $p_\mathit{th}$, the \textit{fitness value} of $x$ given $\delta$ and $p_\mathit{th}$, denoted by $\mathit{fitness}(x, \delta, p_\mathit{th})$, is defined as follows:
\begin{equation}
\mathit{fitness}(x, \delta, p_\mathit{th}) = \frac{\mathit{diff}(p_{x, \delta}, p_\mathit{th})}{\max(p_\mathit{th}, (1-p_\mathit{th}))}
\end{equation}
where the denominator is a normalisation factor, making the range of the fitness value between 0 and 1.
\end{definition}
In other words, we compute the fitness value of an input $x$ using the difference between $p_{x, \delta}$ (i.e., the proportion of unsafe inputs in the neighborhood of $x$ with the radius of $\delta$) and $p_\mathit{th}$ (i.e., the probability threshold).

Note that the fitness function is meant to be minimized and decreases as the difference between $p_\mathit{th}$ and $p_{x, \delta}$ decreases. 
The fitness function also takes the number of observations (i.e., evaluated inputs) within $N(x, \delta)$ into account, as the size of $\mathit{CI}(p_{x, \delta})$ (i.e., the confidence interval of $p_{x, \delta}$) decreases when the number of observations in the neighborhood increases, thereby also decreasing the value of the fitness function. 
A sparsely populated neighborhood therefore tends to yield high fitness values, which is what we would expect as $p_{x, \delta}$ in such neighborhoods comes with much uncertainty.

To better illustrate how the boundary-seeking fitness function distinguishes between inputs based on their proximity to the boundary,
let us consider an input space $X$ and two inputs $x_1\in X$ and $x_2\in X$ where $\mathit{CI}(p_{x_1, \delta}) = 0.1 \pm 0.05$ and $\mathit{CI}(p_{x_2, \delta}) = 0.5 \pm 0.1$ for a small $\delta$.
This means that the proportions of unsafe inputs around $x_1$ and $x_2$ are estimated as $0.1 \pm 0.05$ and $0.5 \pm 0.1$, respectively.
If we consider the boundary of a 5\%-unsafe region (i.e., $p_\mathit{th} = 0.05$), we can say that $x_1$ is closer to the boundary than $x_2$ since the proportion of unsafe inputs around $x_1$ is up to 15\% while that around $x_2$ is up to 60\%. 
This is exactly captured by the fitness function since $\mathit{diff}(p_{x_1, \delta}, 0.05) = 0.1$ and $\mathit{diff}(p_{x_2, \delta}, 0.05) = 0.55$, thus yielding $\mathit{fitness}(x_1, \delta, 0.05) = \frac{0.1}{0.95} = 0.105$ and $\mathit{fitness}(x_2, \delta, 0.05) = \frac{0.55}{0.95} =  0.579$, showing that $x_1$ is closer to the boundary than $x_2$.

\subsection{MLC Systemic Hazard Envelope (\MLCSHE) Algorithm}
\label{sec:algorithm}
Based on the representations of scenarios and MLC behaviors described in Section~\ref{sec:reps} and the boundary-seeking fitness function described in Section~\ref{sec:fitnessFunc}, this section proposes a novel algorithm, \emph{MLC Systemic Hazard Envelope (\MLCSHE)} [/\textipa{mIlS}/], based on CCEA as described at the beginning of Section~\ref{sec:coev}.

Algorithm~\ref{alg:MLCSHE} shows the pseudocode of \MLCSHE. It takes as input
a population size $n$,
a minimum number of joint fitness assessments per individual $k$, 
a threshold probability $p_\mathit{th}$ to define a probabilistic unsafe region, 
a threshold distance $d_a$ to ensure the diversity of individuals and complete solutions in archives, 
a maximum population archive size $l$, 
a distance threshold $d_\mathit{th}$ to filter the complete solutions that are distinct enough,
and a boundary fitness threshold $t_b$ to filter complete solutions close enough to the boundary; 
it returns an archive $A_b$ of distinct complete solutions, with the pairwise distance of more than $d_\mathit{th}$, whose fitness values are less than $t_b$ (i.e., close to the boundary of a $p_\mathit{th}$-unsafe region), 
while $k$ and $l$ are parameters to control the algorithm's search behavior (detailed below).
\MLCSHE in essence is a CCEA that uses population archives as described in Section~\ref{sec:background}. 
However, it is different from other similar methods as its goal is to return a set of complete solutions satisfying certain properties (i.e., close to the boundary) rather than returning a single-best complete solution.

\begin{algorithm}
\SetKwInOut{Input}{Input}
\SetKwInOut{Output}{Output}

\Input{
Population Size $n$\\
Minimum Number of Fitness Assessments per Individual $k$\\
Threshold Probability $p_\mathit{th}$\\
Distance Threshold for Population Archives $d_a$\\
Maximum Size of Population Archive $l$\\
Distance Threshold for Post-processing $d_\mathit{th}$\\
Boundary Fitness Threshold for Post-processing $t_b$
}
\Output{Archive of Distinct Boundary Complete Solutions $A_b$}

\BlankLine
Population of MLC Output Sequences $P_O \gets$ \textit{initPopulation}($n$)~\label{alg:MLCSHE:po}\\
Population of Scenarios $P_S \gets$ \textit{initPopulation}($n$)~\label{alg:MLCSHE:ps}\\
Archive of MLC Output Sequences $A_O \gets P_O$~\label{alg:MLCSHE:ao}\\
Archive of Scenarios $A_S \gets P_S$~\label{alg:MLCSHE:as}\\
Archive of Complete Solutions $A_c \gets \emptyset$~\label{alg:MLCSHE:ac}\\

\While{$\mathit{not (stopping\_condition)}$}{\label{alg:MLCSHE:while} 
	$P_O, P_S, A_c \gets \textbf{\textit{assessFitness}}\big(P_O, P_S, A_O, A_S, k, A_c, p_\mathit{th}\big)$ \label{alg:MLCSHE:cs}\\
    $A_O \gets \textbf{\textit{updatePopulationArchive}}(P_O, l, d_a)$ \label{alg:MLCSHE:updateao}\\
    $A_S \gets \textbf{\textit{updatePopulationArchive}}(P_S, l, d_a)$ \label{alg:MLCSHE:updateas}\\
    
    $P_O \gets \textit{Breed}(P_O)\cup A_O$ \label{alg:MLCSHE:breedpo}\\
    $P_S \gets \textit{Breed}(P_S)\cup A_S$ \label{alg:MLCSHE:breedps}\\
}

Archive of Complete Solutions $A_b \gets \textit{postProcess}(A_c, d_\mathit{th}, t_b) $\label{alg:MLCSHE:postprocess}

\textbf{return} $A_b$\label{alg:MLCSHE:return}\\
    \caption{MLC Hazard Envelope Search algorithm (\MLCSHE)}
\label{alg:MLCSHE}
\end{algorithm}

The algorithm first randomly initializes 
the population of MLC Output sequences $P_O$ (line~\ref{alg:MLCSHE:po}),
the population of scenarios $P_S$ (line~\ref{alg:MLCSHE:ps}),
and their population archives, $A_O$ (line~\ref{alg:MLCSHE:ao}) and $A_S$ (line~\ref{alg:MLCSHE:as}), respectively. 
The algorithm also initializes the archive of complete solutions $A_c$ as an empty set (line~\ref{alg:MLCSHE:ac}).
The algorithm then co-evolves $P_O$ and $P_S$ using $A_O$ and $A_S$, until the $\mathit{stopping\_condition}$ is met (line~\ref{alg:MLCSHE:while}), such that it guides them towards the complete solutions that are close to the boundary of a $p_\mathit{th}$-unsafe region (lines~\ref{alg:MLCSHE:while}--\ref{alg:MLCSHE:breedps}). 
During the co-evolution, the algorithm repeats the following three steps:
\begin{inparaenum}
    \item assess the fitness values of individuals in both $P_O$ and $P_S$ and update $A_c$ to include complete solutions with their joint fitness values evaluated by the simulator (using function \textbf{\textit{assessFitness}} at line~\ref{alg:MLCSHE:cs}, described in detail in Algorithm~\ref{alg:FA});
    \item update $A_O$ and $A_S$ based on the individual fitness values, $d$, and $l$ (using function \textbf{\textit{updatePopulationArchive}} at lines \ref{alg:MLCSHE:updateao}--\ref{alg:MLCSHE:updateas}, described in detail in Algorithm~\ref{alg:UA}); and
    \item evolve $P_O$ and $P_S$ (using the function \textit{breed} detailed at the end of Section~\ref{sec:algorithm}), and merging them with $A_O$ and $A_S$, respectively, to make up the next generation of $P_O$ and $P_S$ (lines~\ref{alg:MLCSHE:breedpo}--\ref{alg:MLCSHE:breedps}).
\end{inparaenum}
After the co-evolution, the algorithm creates a set of complete solutions $A_b$ from $A_c$ such that 
the distance between two arbitrary, complete solutions in $A_b$ is at least $d_\mathit{th}$ 
and the fitness value of every complete solution in $A_b$ is less than $t_b$ (using function \textit{postProcess} at line~\ref{alg:MLCSHE:postprocess}).
The algorithm ends by returning $A_b$ (line \ref{alg:MLCSHE:return}).

\subsubsection{Fitness Assessment}\label{sec:alg:fitAssess}
The function \textbf{\textit{assessFitness}} is to first calculate the joint fitness values of complete solutions, generated by joining the individuals in $P_O$ and $P_S$ (with higher priorities to the individuals in $A_O$ and $A_S$, respectively) such that each individual is joined at least $k$ times, using the simulator.
In other words, the fitness of each individual is assessed based on at least $k$ collaborators to avoid inaccurately estimating the individual fitness (see Section~\ref{sec:background} for more details about collaborators).
To reduce the number of computationally intensive simulations, complete solutions that are the same as the ones in $A_C$ (i.e., generated in the previous generations) are not simulated again. 
Then, the function assesses the fitness value of each individual using the joint fitness values of the complete solutions that contain the individual. 

Specifically, Algorithm~\ref{alg:FA} shows the pseudocode of \textbf{\textit{assessFitness}}.
It takes as input
the population of MLC output sequences $P_O$, 
the population of scenarios $P_S$, 
the population archive of MLC output sequences $A_O$,
the population archive of scenarios $A_S$,
the minimum number of fitness assessments per individual $k$,
the archive of previously evaluated complete solutions $A_C$,
the neighborhood radius $\delta$,
and the threshold probability $p_\mathit{th}$;
it then returns
$P_O$ and $P_S$ updated to include individual fitness values, and
$A_C$ updated to include newly generated complete solutions and their joint fitness values.

\begin{algorithm}
\SetKwInOut{Input}{Input}
\SetKwInOut{Output}{Output}

\Input{
Population of MLC Output Sequences $P_O$\\
Population of Scenarios $P_S$\\
Archive of MLC Output Sequences $A_O$\\
Archive of Scenarios $A_S$\\
Minimum Number of Fitness Assessments per Individual $k$\\
Archive of Complete Solutions $A_c$\\
Complete Solutions Pairwise Distance Matrix $D_c$\\
Neighborhood Radius $\delta$\\
Threshold Probability $p_\mathit{th}$}
\Output{
Updated Population of MLC Output Sequences $P_O$\\
Updated Population of Scenarios $P_S$\\
Updated Archive of Complete Solutions $A_c$}

\BlankLine
Set of Populations $PS \gets \{ P_O, P_S \}$\label{alg:FA:ps}\\

Set of Complete Solutions $C \gets \textit{collaborate}(P_O, P_S, A_O, A_S, k)$\label{alg:FA:c}

\ForEach{Complete Solution $c \in C$}{\label{alg:FA:safetyEval}
    \If{$c \notin A_c$}{\label{alg:FA:if}
        $c.\textit{isUnsafe} \gets \textit{simulate(c)}$\label{alg:FA:sim}\\
        $A_c \gets A_c \cup \{c\}$\label{alg:FA:update_ac}
    }
}

\ForEach{Complete Solution $c \in A_c$}{\label{alg:FA:jfeval}
    $c.\textit{fitness} \gets \textit{computeBoundaryFitness}(c, A_c, \delta, p_\mathit{th})$\label{alg:FA:computeJFV}
}

\ForEach{Population $P \in PS$}{\label{alg:FA:foreachp}
     \ForEach{Individual $i \in P$}{\label{alg:FA:foreachi}
        $i.\textit{fitness} \gets \textit{assessIndividualFitness}(i, A_c)$\label{alg:FA:indfit}
    }
}
\textbf{return} $P_O$, $P_S$, $A_c$\label{alg:FA:return}
\caption{\textbf{\textit{assessFitness}}}
\label{alg:FA}
\end{algorithm}

The algorithm begins by initializing a set of populations $PS$ as $\{P_S, P_O\}$ (line~\ref{alg:FA:ps}). 
It also initializes a set of complete solutions $C$ by selecting and collaborating individuals from $P_S$ and $P_O$ using the \textit{collaborate} function (line~\ref{alg:FA:c}). 
This function first makes every individual of $P_S$ and $P_O$ collaborate with every individual of $A_O$ and $A_S$, respectively, and if the number of collaborations for each individual is less than $k$ (i.e., when the size of $A_O$ and $A_S$ are less than $k$, where $k\geq 1$), randomly selected individuals of $P_O \setminus A_O$ and $P_S \setminus A_S$ are used in addition to $A_O$ and $A_S$, respectively, to ensure a minimum of $k$ collaborations for each individual
\footnote{Note that a high value of $k$ might add significant computational cost to the search since it tends to exponentially increase the number of joint fitness evaluations per individual.}.
Then, for each complete solution $c\in C$ (line~\ref{alg:FA:safetyEval}), if $c\notin A_C$, i.e., $c$ has not been previously evaluated (line~\ref{alg:FA:if}), the algorithm evaluates $c$ using the high-fidelity simulator to identify if $c$ is unsafe (line~\ref{alg:FA:sim}) and adds $c$ with its evaluated result into $A_c$ (line~\ref{alg:FA:update_ac}).
Once $A_c$ is updated using $C$, for each complete solution $c \in A_c$ (line~\ref{alg:FA:jfeval}), the algorithm computes its joint fitness value (i.e., the boundary-seeking fitness value) using $A_c$, $\delta$, and $p_\mathit{th}$ by calculating the proportion of unsafe complete solutions in the neighborhood of $c$ and its difference from the threshold probability as described in Section~\ref{sec:fitnessFunc} (line~\ref{alg:FA:computeJFV}).
Although computing the neighborhood of $c$ requires many distance computations, we can significantly reduce the computations by reusing the distances among the complete solutions that were originally in the input $A_c$.
For each individual $P \in PS$ and for each $i \in P$ (lines~\ref{alg:FA:foreachp}--\ref{alg:FA:foreachi}), the algorithm sets the minimum (i.e., the best since we aim to minimize fitness values) joint fitness value of the complete solutions involving $i$ as the individual fitness of $i$ (line~\ref{alg:FA:indfit}).
An elitist individual fitness assessment strategy (i.e., selecting the best fitness value), is consistent with reported experimental studies~\cite{ma2019survey, Luke2013Metaheuristics} as well as our preliminary evaluation results on a widely used benchmark problem known as the \emph{MTQ} (\emph{Maximum of Two Quadratics})~\cite{bucci2005identifying}.
The algorithm ends by returning the updated $P_O$, $P_S$, and $A_c$ (line~\ref{alg:FA:return}).

\subsubsection{Update Population Archive}
\label{sec:updateArchive}
The function \textbf{\textit{updatePopulationArchive}} updates the population archives $A_O$ and $A_S$, for the next generation. They play a key role in guiding the search algorithm since every other individual has to form a complete solution with them, whose joint fitness will be assessed afterwards.

There are many ways to update the population archive such as the ones proposed in \textit{iCCEA} and \textit{pCCEA}~\cite{ma2019survey}. However, as mentioned in Section~\ref{sec:background}, they can be inefficient due to additional fitness evaluations for updating population archives, making them impractical for our problem involving computationally expensive simulations for fitness evaluation. 
Instead, we can consider more efficient archive update strategies as follows: selecting individuals with the best fitness values (\textit{Best}), selecting the best individual plus random individuals (\textit{Best$+$Random}), or randomly selecting individuals (\textit{Random})~\cite{ma2019survey}. 
Our preliminary evaluation results on the \emph{MTQ} problem showed that both \textit{Best} and \textit{Best$+$Random} work similarly well for updating population archives in \MLCSHE. To ensure the diversity of individuals in each population archive and maximize exploration, we choose \textit{Best$+$Random} with a similarity threshold (i.e., the distance threshold $d_\mathit{th}$) that filters out individuals deemed too similar to be included in a population archive. The pseudocode for updating a population archive is provided in Algorithm~\ref{alg:UA}.

The algorithm takes as input
a target population $P$,
a maximum size of a population archive $l$,
and a threshold distance (i.e., the minimum distance between two arbitrary individuals in a population archive) $d_\mathit{th}$;
it returns a population archive $A_P$ of $P$ such that $|A_P| \leq l$ and $d(i, j) \ge d_\mathit{th}$, based on the distance function $d$ as described in Section~\ref{sec:fitnessFunc}, for all $i, j\in A_P$ if $i \neq j$. 

\begin{algorithm}
\SetKwInOut{Input}{Input}
\SetKwInOut{Output}{Output}

\Input{Population $P$\\
Maximum Size of Population Archive $l$\\
Threshold Distance $d_\mathit{th}$\\
}
\Output{Population Archive $A_P$}

\BlankLine

Archive $A_P \gets \{\textit{popBestFitnessIndividual}(P)\}$\label{alg:UA:a}\\

\While{$|A_P| < l$ \textbf{and} $|P| > 0$ }{
    Individual $i \gets \textit{randomPop}(P)$\label{alg:UA:pop}\\
    \If{$\textit{isDistinct}(i, A_P, d_\mathit{th})$}{\label{alg:UA:ifDistinct}
        $A_P \gets A_P \cup \{i\}$\label{alg:UA:addInd}
    }
}

\textbf{return} $A_P$\label{alg:UA:return}
\caption{\textbf{\textit{updatePopulationArchive}}}
\label{alg:UA}
\end{algorithm}

The algorithm starts by initializing a population archive $A_P$ for $P$ using the individual with the best fitness value among all the individuals in $P$ (using function \textit{popBestFitnessIndividual} at line~\ref{alg:UA:a}).
While $|A_P| < l$ or $|P| > 0$, the algorithm iteratively pops a random individual $i$ from $P$ (line~\ref{alg:UA:pop}) and add $i$ into $A_P$ (line~\ref{alg:UA:addInd}) if $i$ is distinct from all individuals in $A_P$ based on the distance threshold of $d_\mathit{th}$ (line~\ref{alg:UA:ifDistinct}).
The algorithm ends by returning $A_P$ (line~\ref{alg:UA:return}).

\subsubsection{Evolution}\label{sec:evo-ops}
As illustrated in Algorithm~\ref{alg:MLCSHE}, after updating $A_O$ and $A_S$, $P_O$ and $P_S$ undergo evolution to generate their next generation using a \emph{breed} operation, which entails three main steps:
\begin{enumerate}
    \item \emph{Selection.} \MLCSHE selects the candidate individuals for breeding via the standard \textit{tournament selection} technique, i.e., the most widely used selection technique for evolutionary algorithms~\cite{Whitley2019}. 
    It is simple yet effective since it only requires the rank ordering of individuals in terms of their fitness values.
    
    \item\emph{Crossover.} Selected individuals of each population act as parents to create offspring individuals using a crossover operation~\cite{Whitley2019}. For our problem, the widely used \textit{uniform crossover} technique~\cite{Luke2013Metaheuristics,Whitley2019} is used, since there is no preference for specific points in individuals as crossover points. 
    
    \item\emph{Mutation.} Finally, offspring individuals are mutated via the introduction of stochastic noise~\cite{Whitley2019}. Since the individuals are heterogeneous vectors with both float and integer values, the standard \textit{Gaussian} and \textit{integer randomization} mutation techniques are applied to individual elements, respectively~\cite{Luke2013Metaheuristics}. 
    Through these mutations, all valid individuals can be considered during the search, making it possible to find the global optimum.
    There is a chance a mutated individual might be invalid. For example, the x-coordinates ($x_{min}$ and $x_{max}$) of a bounding box used to define a detected obstacle can be out of the camera frame width bounds (from 0 to 800 pixels).
    Such invalid cases are handled by a simple \textit{repair} function in our implementation, which is available in the replication package (see Section~\ref{sec:data-availability}).
\end{enumerate}

We want to note that there are hyperparameter values for selection, crossover, and mutation (e.g., the tournament size, crossover and mutation rates) that could affect breeding performance. 
More details on tuning hyperparameter values in our evaluation is provided in Section~\ref{sec:rq1-method}.

\section{Evaluation}\label{sec:evaluation}

In this section, we report on the empirical evaluation of \MLCSHE when applied to an open-source MLAS. Specifically, we provide answers for the following research questions:
\begin{description}[nolistsep]
    \item[RQ1 (Effectiveness)] How \emph{effectively} can \MLCSHE find the MLAS hazard envelop boundary compared to baseline boundary search approaches?
    \item[RQ2 (Efficiency)] How \emph{efficiently} can \MLCSHE find the MLAS hazard envelop boundary compared to baseline boundary search approaches?
\end{description}

To answer RQ1, we investigate \textit{how many} complete solutions (i.e., combinations of scenarios and MLC behaviors) that are close to the boundaries are found by different boundary search approaches, including \MLCSHE, given a same time budget. To answer RQ2, we investigate \textit{how quickly} complete solutions that are close to the boundaries are found by different boundary search approaches.
The results of RQ1 and RQ2 may depend on the distance threshold between complete solutions and the boundaries\footnote{Recall that the fitness of a complete solution is defined based on its distance from the hazard boundary. Throughout the rest of the section boundary distance threshold and fitness threshold are used interchangeably.} ($d_\mathit{th}$ and $t_b$ in algorithm~\ref{alg:MLCSHE}).
Thus, we also consider the effects of the distance thresholds while answering RQ1 and RQ2.

\subsection{Evaluation Subjects}
\label{sec:eval-subjects}

We use Pylot~\cite{Gog2021Pylot}, one of the highest ranking component-based AV on the CARLA Autonomous Driving Leaderboard~\cite{CarlaLeadeboard}, at the time of the evaluation. The leaderboard evaluates AV according to 11 metrics that are designed to assess safe driving performance such as collision, red light infractions and route completion. Pylot's high performance on the leaderboard makes it a good candidate to consider as an evaluation case study. Furthermore, Pylot is one of the only high-ranking AV that is open-source, has been deployed on a real-life vehicle~\cite{Gog2021Pylot}.

We also use CARLA~\cite{Dosovitskiy2017Carla}, a high-fidelity open-source AV simulator. CARLA allows us to control various static and dynamic elements in driving environments. Based on the controllable elements, following a previous study using CARLA~\cite{haq2022samota}, we consider the following seven scenario elements: road curve and length, start and end points on maps, the density of pedestrians, time of day, and weather condition. The detailed explanation for the scenario elements and their values (ranges) are available in the supporting material (see Section~\ref{sec:data-availability}).

For the ML component under test in Pylot, we target a DNN-based obstacle detection module. 
The obstacle detection module takes digital images of the front-facing camera and detects obstacles in the images in terms of their location and size (captured as a bounding box), their type, and the uncertainty associated with the predicted label. The output of the obstacle detection module is then used by an obstacle prediction module that predicts the trajectories of the detected obstacles for future timestamps, followed by planning and control modules that generate driving commands considering the obstacles' predicted trajectories. 
Thus, we let the boundary search methods manipulate the parameters that define the output sequence of the target MLC (i.e., the object detection module) during the execution of a simulation. Specifically, for each obstacle, there are 11 parameters: the label of the detected obstacles (pedestrian or vehicle), the start and end time the obstacles are detected, and the 2D coordinates (i.e., $x_{min}$, $x_{max}$, $y_{min}$ and $y_{max}$) that define the start and end bounding boxes of the trajectories on the input image.
Additional details regarding the scenario and MLC output are available in the supporting material (see Section~\ref{sec:data-availability}).

The above-mentioned categorical (e.g., road curve, weather condition, and obstacle's label) and numeric (e.g., obstacle's position) parameters defining scenarios and MLC outputs are used to define a distance function \textit{dist} which measures the distance between two complete solutions as discussed in Section~\ref{sec:fitnessFunc}. 
Given that these parameters are heterogeneous, we use a \textit{heterogeneous distance metric}. Specifically, \textit{dist} is defined as the average of the normalized Hamming distance~\cite{2020SciPy-NMeth} of categorical values and the normalized City Block distance~\cite{2020SciPy-NMeth} of numeric values, where the latter are normalized by their maximum range of values that each parameter can take. For instance, the $y$-coordinates of the MLC outputs range from 0 to 600 due to the height of the camera frame, and thus they are divided by 600 to be normalized.
As we have many pairwise distance calculations during the search, we opted for these distance metrics since they are computationally efficient compared to alternatives.
Given its definition above, \textit{dist} ranges between 0 and 1. If $\textit{dist}=1$ between two complete solutions, it means all the categorical values of the two are different, and the differences in all the numeric values of the two are the maximum.

Among various AV safety requirements used in the literature~\cite{haq2021can, lundgren2021safety, riberio2022REslr}, considering the capability of CARLA and the major functionality of our target MLC (i.e., the object detection module), we focus on the following safety requirement: 
\textit{``AV should have a distance no less than $d_{min}$ from the vehicle in front.''}
To detect safety violations, if any, during the simulation of a complete solution (i.e., the combination of a scenario and an MLC behavior), we measure the distance between the ego vehicle and the vehicle in front for each simulation time step.
If the distance is less than $d_{min}$ at any time, the violation is detected and the complete solution is marked as unsafe.
Since our driving scenarios often involve junctions with traffic lights where the vehicles should completely stop for a while, $d_{min}$ should be small enough to avoid false alarms (i.e., incorrectly triggering safety violations) even when the vehicles completely stop. Based on the rule-of-thumb that the driver should be able to see the rear tire of the vehicle in front and a small part of the asphalt when stopping behind a stopped vehicle, we set $d_{min}$ to \SI{1.5}{\meter}.

Due to the execution time of individual simulations in CARLA, which is around five minutes on average, the total computing time for the evaluation is more than \emph{1800 hours} (\emph{75 days}).
To address this issue, we conduct our evaluation on two machines, M1 and M2. 
Machine M1 is a desktop computer with 2.6 GHz Intel i7-10750H CPU, NVidia GeForce RTX 2070 with Max-Q Design GPU (with 8 GB memory), and 32 GB RAM, running Ubuntu 20.04.
Machine M2 is a \textsf{g4dn.2xlarge} node configured as NVIDIA GPU-Optimized AMI (version 22.06.0) in Amazon Elastic Cloud (EC2) with eight virtual cores, NVIDIA T4 GPU (with 16GB memory), and 32 GB RAM, running Ubuntu 20.04.
Specifically, we use M1 for Random Search (RS) and standard Genetic Algorithm (GA), while \MLCSHE is run on M2.
Note that since we keep the number of simulations, as opposed to the execution time, constant over all the experiments, the experiments on M1 and M2 are comparable (see Section~\ref{sec:rq1-method} for details).

\subsection{RQ1: Effectiveness}\label{sec:effectiveness}

\subsubsection{Methodology}\label{sec:rq1-method}

To answer RQ1, we execute \MLCSHE and other comparable methods to generate sets of complete solutions that are close to the boundary and measure their boundary search effectiveness in terms of \textit{Distinct Boundary Solutions (DBS)} capturing the number of distinct complete solutions close to the boundary.
Specifically, given a distinctiveness (distance) threshold $d_\mathit{th}$ (for the distinctiveness of complete solutions) and a boundary closeness (fitness) threshold $t_b$ (for the closeness to the boundary), let $C_V$ be the set of complete solutions generated by a boundary-seeking method $V$, satisfying the following conditions\footnote{$C_V$ is computed via the post-processing function \textit{postProcess} shown in Algorithm~\ref{alg:MLCSHE}.}: 
(1) the pairwise distance between two arbitrary complete solutions in $C_V$ is more than $d_\mathit{th}$ and 
(2) the fitness value of every complete solution in $C_V$ is less than $t_b$.
Then, \textit{DBS} of $V$ is defined as $\textit{DBS}(V) = |C_V|$ (i.e., the size of $C_V$). 
Recall that both distance and fitness values are normalized ($d_\mathit{th}, t_b \in [0, 1]$).

To better understand how \textit{DBS} varies depending on different $d_\mathit{th}$ and $t_b$ thresholds, we vary $d_\mathit{th}$ and $t_b$. 
Specifically, we set $d_\mathit{th}$ to 0.1, 0.2, and 0.3 because it is unrealistic to think that two arbitrary complete solutions are distinctive only if their pairwise distances are more than 30\% of the maximum possible distance\footnote{We actually confirmed that, when $d_\mathit{th} > 0.3$, even \app yields insufficient DBS for safety monitoring. The results for $d_\mathit{th} > 0.3$ are also included in our replication package~\cite{Replication_Package}.}.
We set $t_b$ to 0.01, 0.03, 0.5, 0.1, 0.15, and 0.2 because we are not interested in complete solutions with fitness values above 0.2 (i.e., far from the boundary, such that the normalized difference between the probability threshold $p_\mathit{th}$ and the proportion of unsafe inputs near the complete solutions is above 0.2).

For the other methods to compare with \MLCSHE, as discussed in section~\ref{sec:relatedWork}, we could not find any other work that has been proposed to address the problem targeted by this paper. 
Note that DeepJanus is incomparable to \MLCSHE, as discussed in Section~\ref{sec:relatedWork}, because:
\begin{enumerate*}
    \item its goal is to study an MLC's safety under various conditions, which is different than the goal of this research effort, i.e., finding the conditions under which an MLC's behavior can impact the safety of the system;
    \item the boundary identified by DeepJanus consists of safe-unsafe pairs that can exist in probabilistic safe or unsafe regions.
\end{enumerate*}
Thus, we compare \MLCSHE against two baseline methods, namely \emph{Random Search (RS)} and \emph{standard Genetic Algorithm (GA)}~\cite{Luke2013Metaheuristics}. RS randomly generates complete solutions, and GA evolves complete solutions without considering two separate populations of scenarios and MLC behaviors. In all the methods (including \MLCSHE), the fitness function is the same as defined in Section~\ref{sec:fitnessFunc}. The results of RS will show how difficult the search problem is. Furthermore, the comparison between \MLCSHE and GA will show how effective our CCEA-based method is compared to a standard search method. 

For all methods, we set the total number of simulations as the search budget to 1,300 (i.e., around 2.5 days to run with two parallel simulations per run), which was a large enough number to see the convergence of the effectiveness metrics on our preliminary evaluation. 
Since most of the execution cost is dedicated to running simulations, the computation budget of the experiments is mainly determined by the number of simulations. Thus, we use the total number of simulations as the search budget. 
Note that, for \MLCSHE and GA, the actual number of simulations could be slightly more than the predefined total number since population-based method check if the search budget is exhausted only after the completion of one generation. 
In addition to the search budget, to ensure the comparability, we set the same \emph{boundary threshold probability} (i.e., $p_{th}$) and the same maximum number of obstacle trajectories per $\mathit{mlco}$ to 0.1 and 2, respectively, for all the methods. This makes a complete solution to have $7$ ($scenario$) $+$ $2 \times 11$ ($mlco$) $= 29$ dimensions.

\MLCSHE and GA have additional hyperparameters.
For GA, we used recommended values in~\cite{Mirjalili2019GA}; the population size, the mutation rate, and the crossover rate are set to 60, 0.01, and 0.85, respectively.
However, since there are no suggested values for CCEAs, for which there is much less experience, we decided to tune them on two benchmark problems, namely MTQ and Onemax, that are widely used in evaluating CCEAs~\cite{ma2019survey}. 
As a result, we used the following hyperparameters for \MLCSHE:  population size = 10, maximum population archive size = 3,  mutation rate = 1.0, and crossover rate = 0.5. The reason for the high mutation rate is to compensate for the individuals in the population archives that are directly passed to the next generation without mutation and crossover in CCEAs.
Similarly, regarding the distinctiveness threshold for population archives ($d_a$) in \MLCSHE, we set it to 0.4 based on the two benchmark results.

To account for the randomness of the search-based methods, we repeat the experiments for each method 10 times.
To evaluate the statistical significance of the difference in effectiveness metrics of different search methods, we use the Mann-Whitney U test~\cite{mann1947test}.
To measure the effect size of the differences, we measure Vargha and Delaney's $\hat{A}_{AB}$, where $0 \leq\hat{A}_{AB} \leq 1$~\cite{vargha2000effect}. Typically, the value of $\hat{A}_{AB}$ indicates a small, medium, and large difference (effect size) between populations $A$ and $B$ when it is higher than 0.56, 0.64, and 0.71, respectively.

\subsubsection{Results}\label{sec:rq1-results}

Table~\ref{tab:rq1-dbs} reports the \textit{DBS} achieved by \MLCSHE, RS and GA over 10 runs at various distinctiveness (distance) threshold ($d_\mathit{th}$) and boundary closeness (fitness) threshold ($t_b$) values.

\begin{table*}
  \centering
  \small
  \caption{\textit{DBS} values for different search methods at different values of $t_b$ and $d_\mathit{th}$.}
  \label{tab:rq1-dbs}
  \resizebox{0.75\textwidth}{!}{
  \begin{tabular}{cccccccc}
  \toprule
    &&\multicolumn{6}{c}{Average \textit{DBS}$\pm 0.5\times CI_{0.95}$}\\
    \cmidrule(lr){3-8}
    &&$t_b=0.01$&$t_b=0.03$&$t_b=0.05$&$t_b=0.10$&$t_b=0.15$&$t_b=0.20$
    \\
    \midrule
    \multirow{3}{2em}{$d_\mathit{th}=0.1$}&RS&$0.0$&$0.0$&$0.0$&$0.5\pm 0.4$&$5.1\pm 2.0$&$19.0\pm 8.8$\\
    &GA&$0.0$&$12.0\pm 6.1$&$37.2\pm 8.9$&$66.7\pm 11.4$&$82.8\pm 10.8$&$94.1\pm 15.4$\\
    &MLCSHE&$0.0$&$0.0$&$39.6\pm 17.2$&$166.8\pm 33.2$&$247.5\pm 29.2$&$315.4\pm 35.1$\\
    \cmidrule(lr){1-2}\cmidrule(lr){3-8}
    \multirow{3}{2em}{$d_\mathit{th}=0.2$}&RS&$0.0$&$0.0$&$0.0$&$0.5\pm 0.4$&$4.6\pm 1.5$&$17.0\pm 7.3$\\
    &GA&$0.0$&$3.8\pm 1.6$&$9.2\pm 2.5$&$18.8\pm 3.3$&$26.0\pm 2.6$&$29.8\pm 2.8$\\
    &MLCSHE&$0.0$&$0.0$&$19.4\pm 6.8$&$57.5\pm 6.7$&$82.0\pm 5.2$&$101.6\pm 6.5$\\
    \cmidrule(lr){1-2}\cmidrule(lr){3-8}
    \multirow{3}{2em}{$d_\mathit{th}=0.3$}&RS&$0.0$&$0.0$&$0.0$&$0.4\pm 0.3$&$2.6\pm 0.6$&$6.7\pm 2.1$\\
    &GA&$0.0$&$1.5\pm 0.7$&$3.7\pm 0.9$&$8.0\pm 1.0$&$11.2\pm 1.2$&$12.3\pm 1.1$\\
    &MLCSHE&$0.0$&$0.0$&$7.9\pm 1.8$&$17.9\pm 1.1$&$23.1\pm 1.7$&$25.6\pm 2.2$\\
    \bottomrule
\end{tabular}
}
\end{table*}

Overall, for all three boundary-seeking methods, \textit{DBS} values increase as boundary closeness threshold ($t_b$) values increase. This is expected since increasing the value of $t_b$ results in more boundary solutions to consider.
Further, \textit{DBS} values drop rapidly as distinctiveness threshold ($d_\mathit{th}$) values increase. This is also expected since identifying complete solutions that are distinct enough with respect to higher $d_\mathit{th}$ values becomes quickly more challenging.
However, when $t_b=0.01$, the \textit{DBS} equals to zero for all methods regardless of $d_\mathit{th}$, meaning that none of the methods found complete solutions closer to the boundary than 0.01. This is simply because the provided search budget (1,300 simulations) is not enough to decrease the fitness values of complete solutions below 0.01 by reducing the size of confidence intervals (i.e., Equation~\ref{eq:confInterval}).

\begin{figure*}
\begin{subfigure}{0.32\textwidth}
  \centering
  \includegraphics[width=\textwidth]{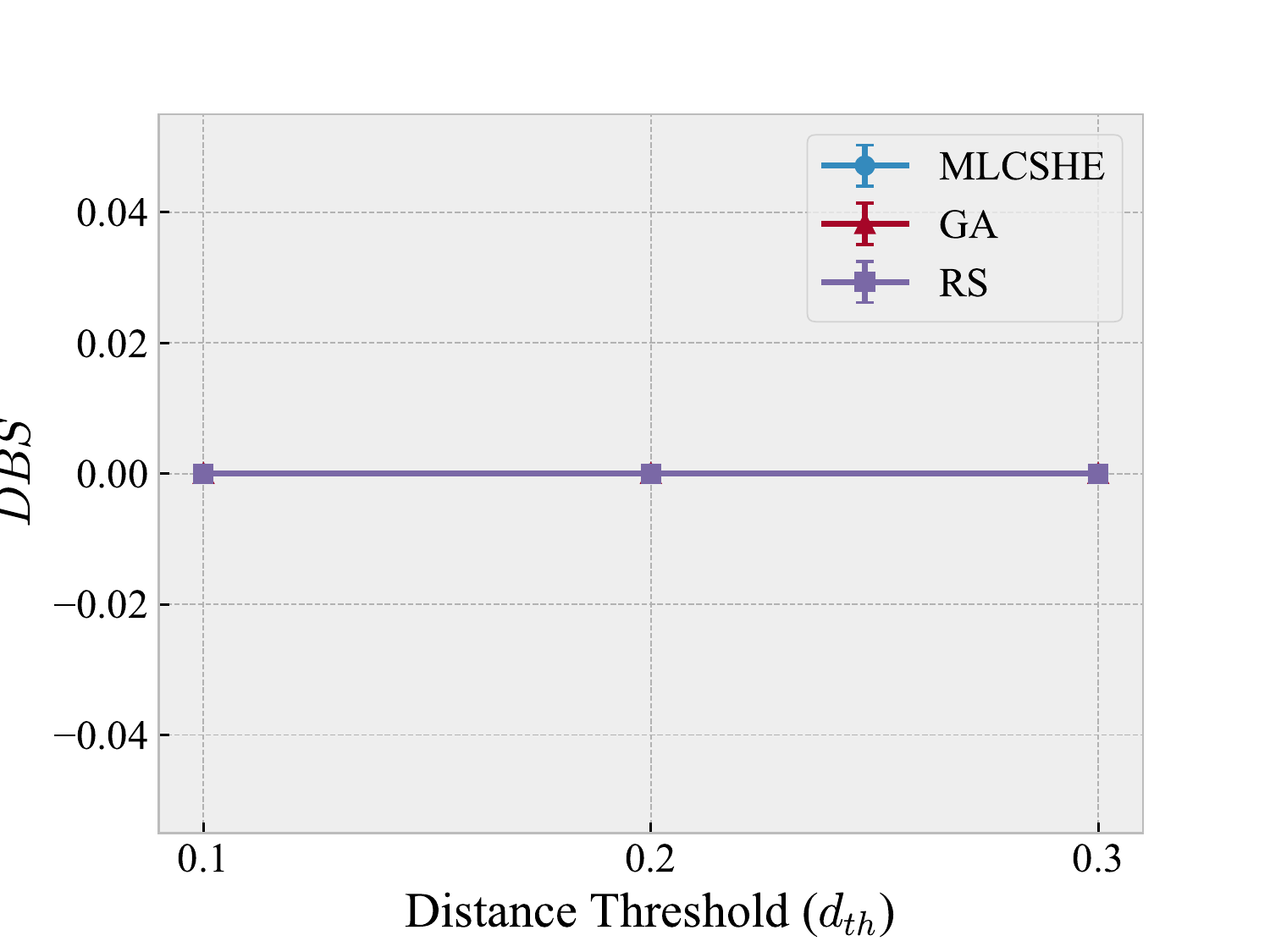}
  \caption{$t_b=0.01$}
  \label{fig:ft-01}
\end{subfigure}
\hfill
\begin{subfigure}{0.32\textwidth}
  \centering
  \includegraphics[width=\textwidth]{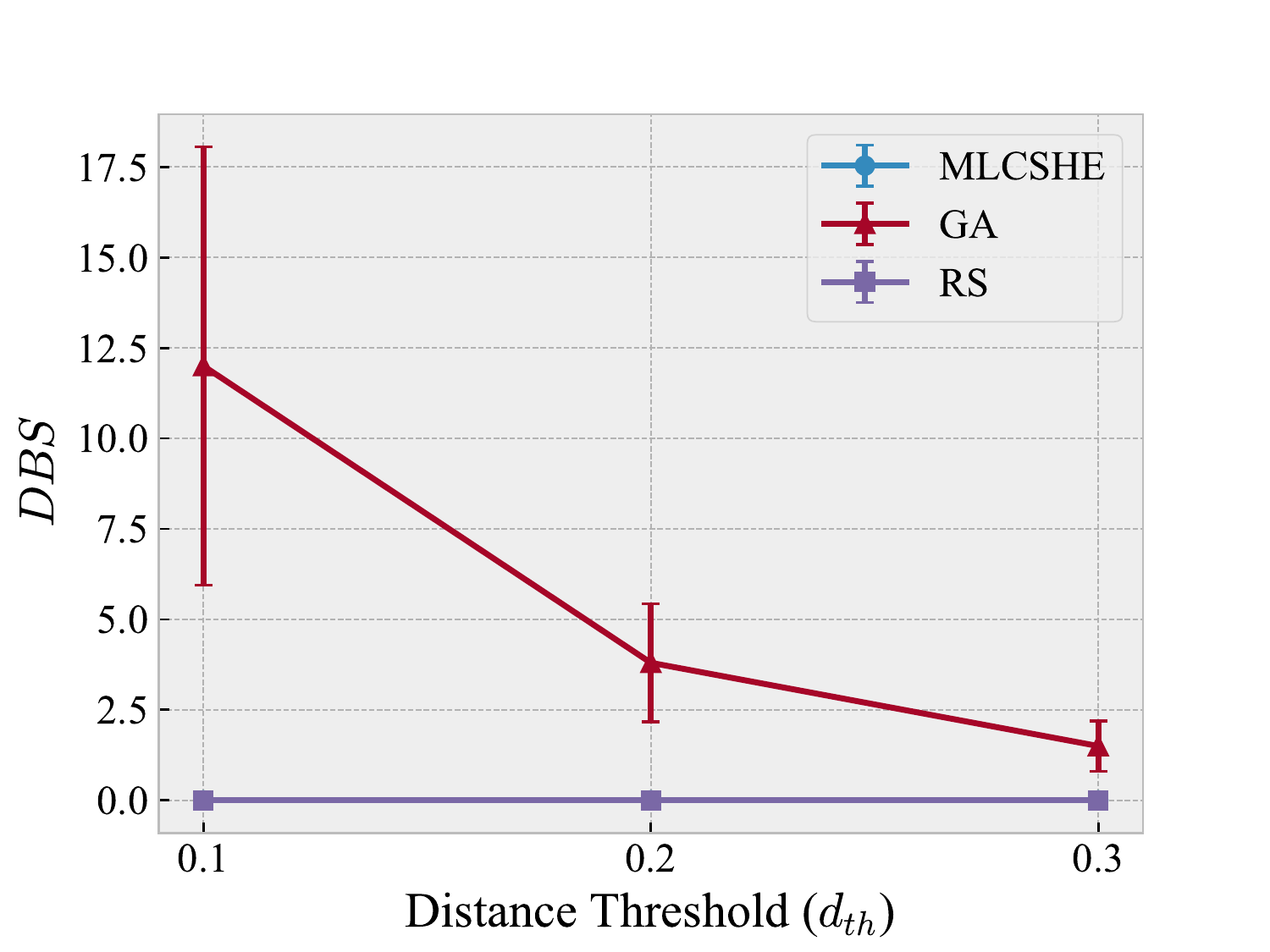}
  \caption{$t_b=0.03$}
  \label{fig:ft-03}
\end{subfigure}
\hfill
\begin{subfigure}{0.32\textwidth}
  \centering
  \includegraphics[width=\textwidth]{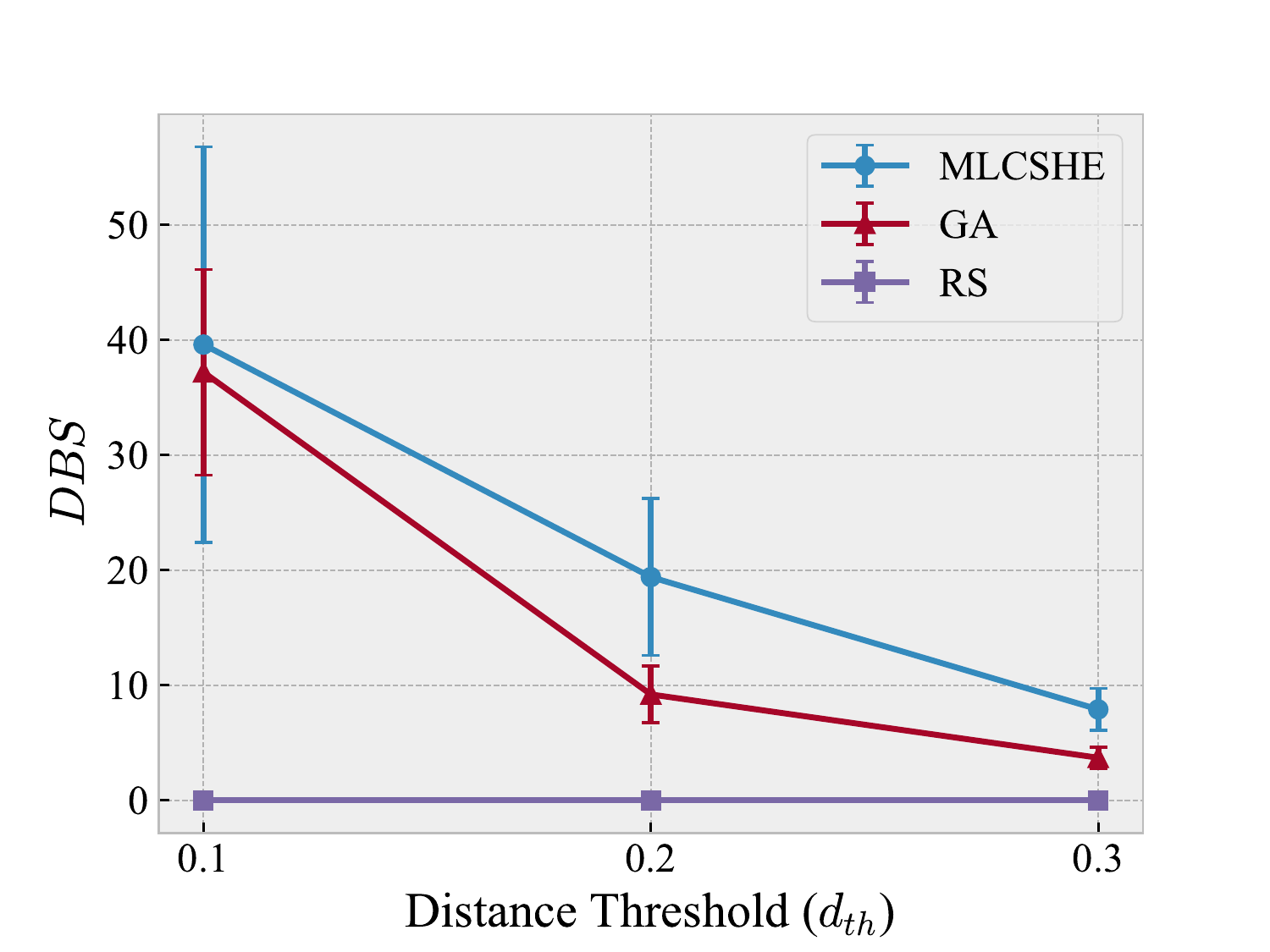}
  \caption{$t_b=0.05$}
  \label{fig:ft-05}
\end{subfigure}
\hfill
\begin{subfigure}{0.32\textwidth}
  \centering
  \includegraphics[width=\textwidth]{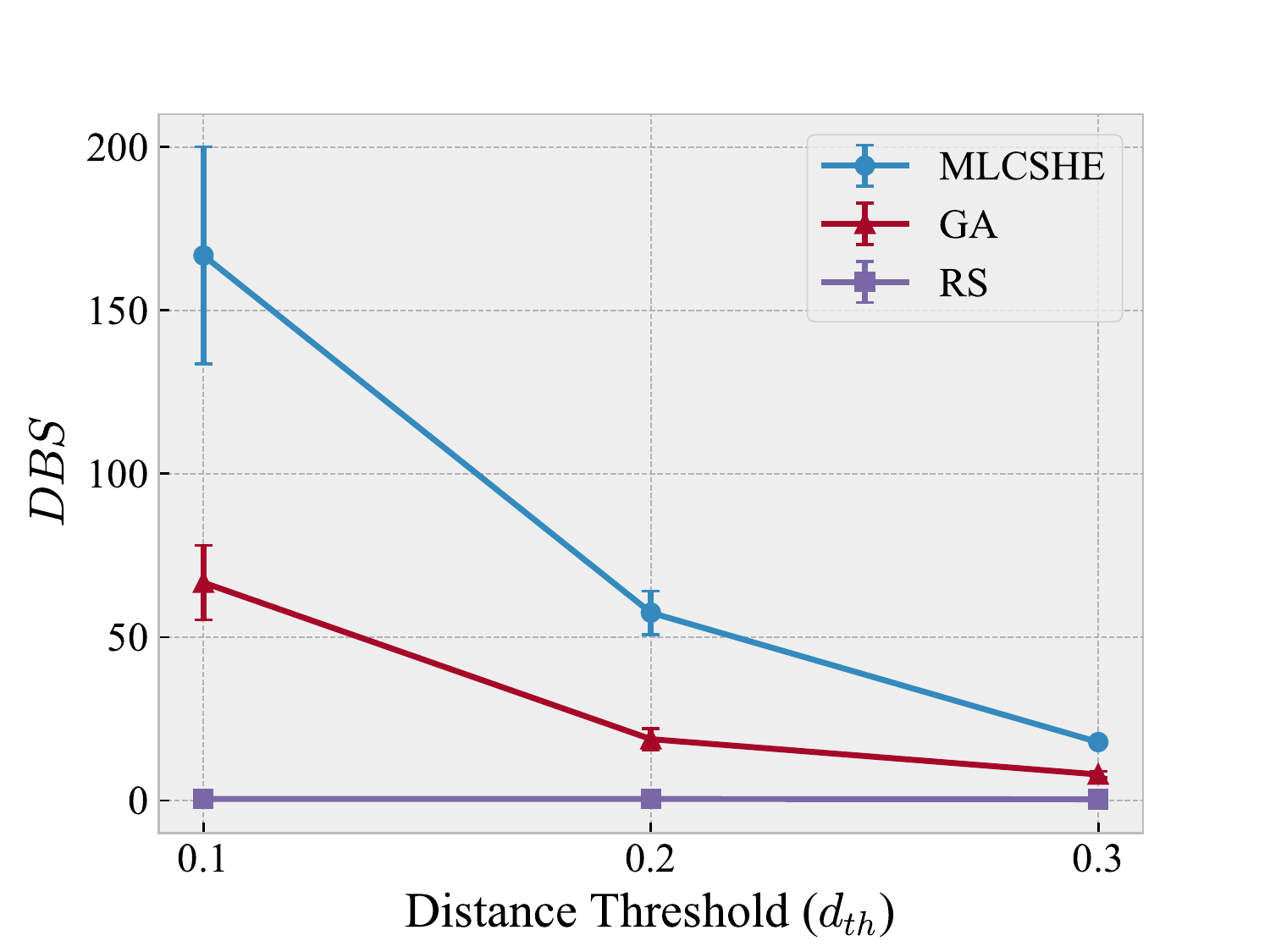}
  \caption{$t_b=0.10$}
  \label{fig:ft-10}
\end{subfigure}
\hfill
\begin{subfigure}{0.32\textwidth}
  \centering
  \includegraphics[width=\textwidth]{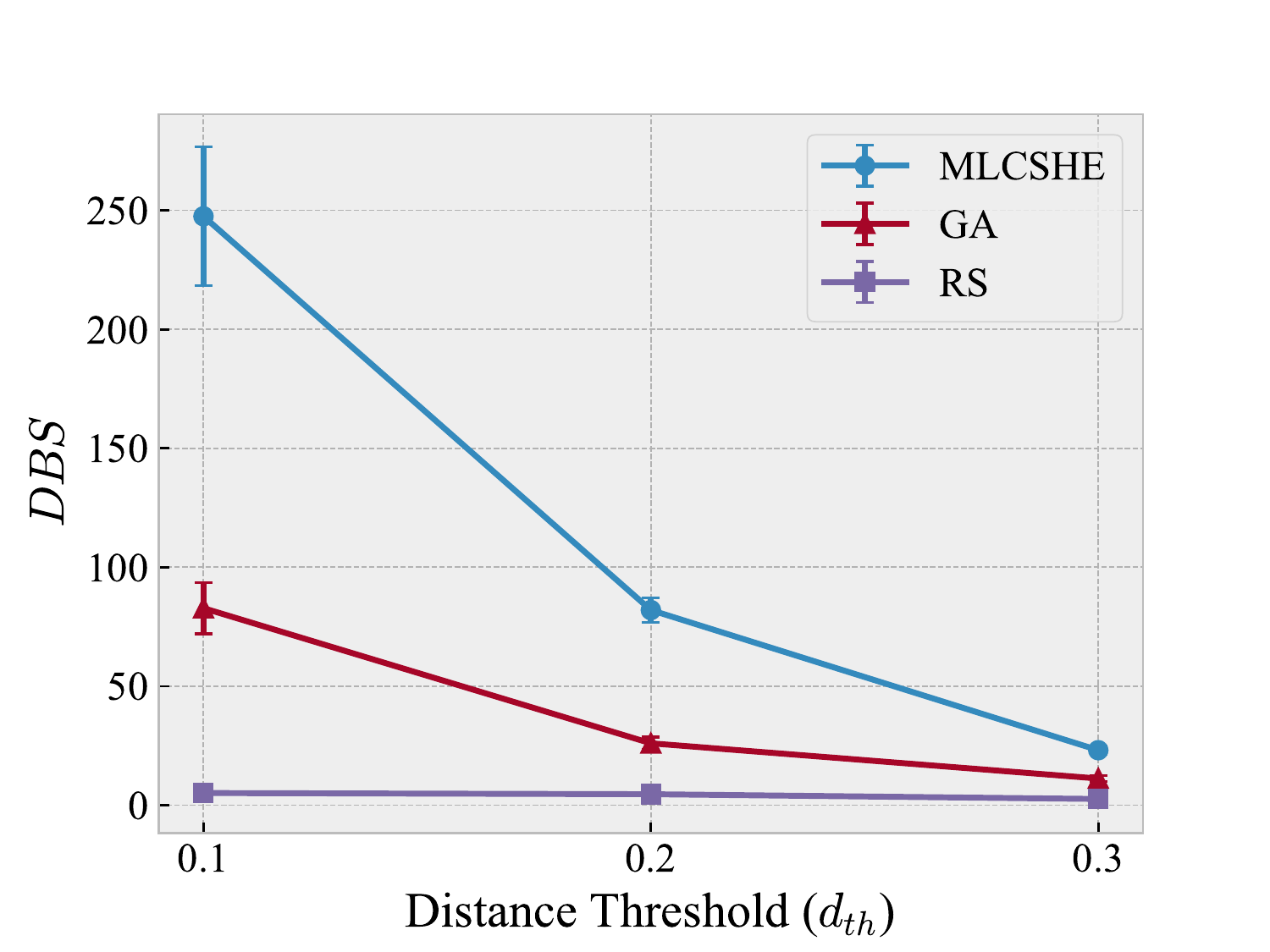}
  \caption{$t_b=0.15$}
  \label{fig:ft-15}
\end{subfigure}
\hfill
\begin{subfigure}{0.32\textwidth}
  \centering
  \includegraphics[width=\textwidth]{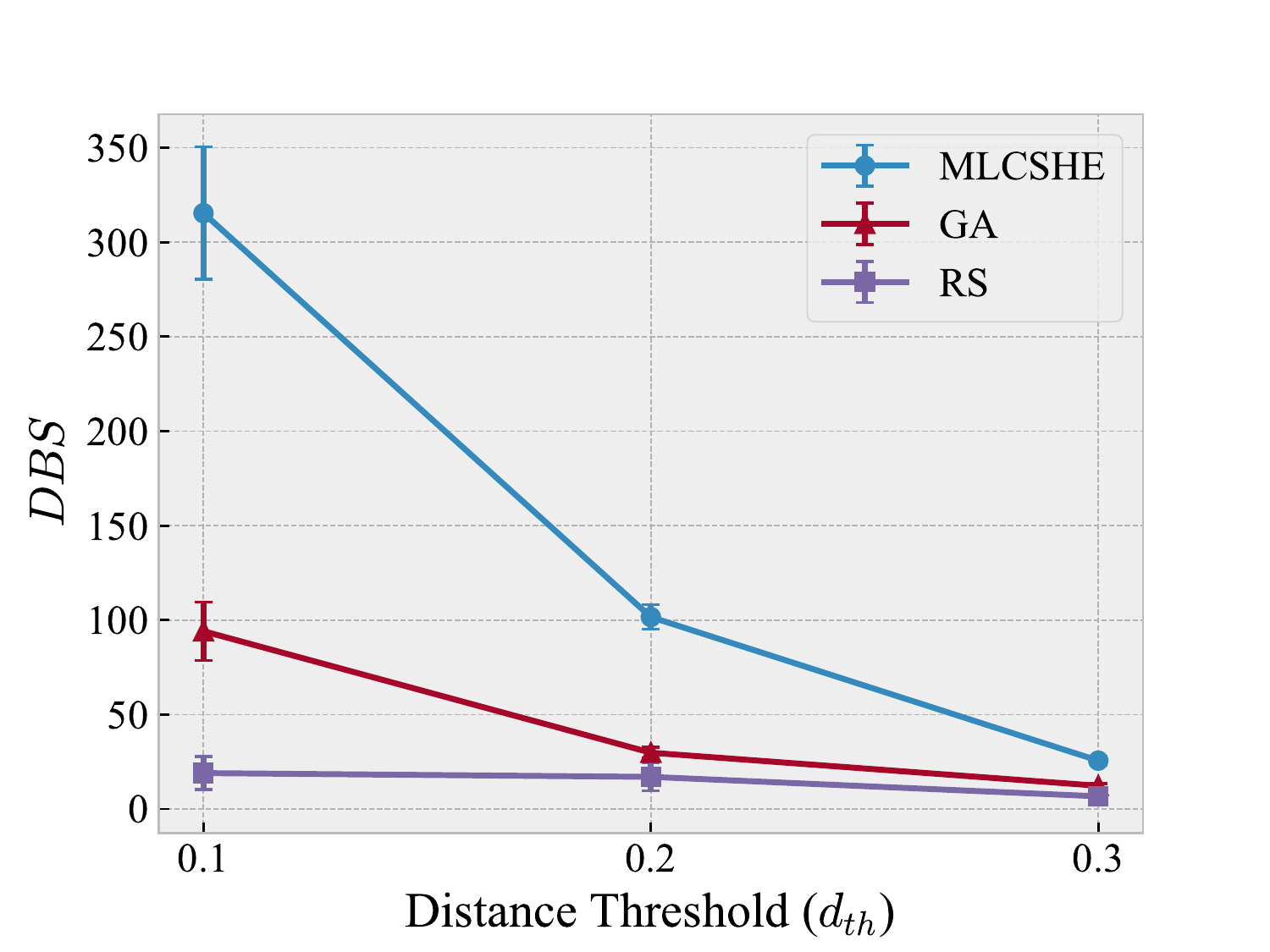}
  \caption{$t_b=0.20$}
  \label{fig:ft-20}
\end{subfigure}
\caption{The relationship between $d_\mathit{th}$ (distinctiveness threshold) and \textit{DBS} (distinct boundary solutions) along with their confidence intervals (shown as error bars) for \MLCSHE, GA, and RS for different $t_b$ (boundary closeness threshold) values.}
\label{fig:dbs-v-dt}
\end{figure*}

\figurename~\ref{fig:dbs-v-dt} depicts how the \textit{DBS} values of the different methods vary with increasing $d_\mathit{th}$ for different $t_b$ values. In each plot, the x-axis is $d_\mathit{th}$ and the y-axis is the average \textit{DBS} over 10 repeats. The \textit{DBS} values for \MLCSHE, GA, and RS are marked with circles, triangles, and squares, respectively. The 95\% confidence intervals for the average \textit{DBS} values are also shown as error bars.

First, RS achieves extremely low \textit{DBS} values when compared to the other two methods in all cases. This implies that the problem of identifying MLAS boundaries is sufficiently challenging for RS not to be able to satisfactorily address it. 

Regarding \app and GA, we can see different patterns depending on different $t_b$ values.
When $t_b = 0.01$, due to the limited search budget provided as already discussed above, \textit{DBS} = 0 for all the methods, meaning that none of the methods find distinct complete solutions near the boundary.
When $t_b = 0.03$, \app once again does not find any complete solutions near the boundary, whereas GA finds some (e.g., $12.0 \pm 6.1$ for $d_\mathit{th}=0.1$), which are likely not applicable for safety monitoring as a very focused and limited part of the hazard boundary is covered by these solutions.
However, when $t_b \geq 0.05$, \app finds more complete solutions near the boundary (e.g., $39.6 \pm 17.2$ for $d_\mathit{th}=0.1$ and $t_b = 0.05$) than GA.
Furthermore, for the same $d_\mathit{th}$ value, the gap between \app and GA increases significantly as $t_b$ increases.

A plausible explanation for these results is that, although GA is better than \app in terms of exploiting a specific region, leading to higher \textit{DBS} values when the boundary closeness threshold is very low compared to the provided simulation budget (i.e., when $t_b = 0.03$), \app successfully uses a cooperative co-evolutionary algorithm which decomposes a high-dimensional problem into two lower-dimensional sub-problems, making the search more effective than GA in terms of identifying distinctive (diverse) complete solutions near the boundary for higher $t_b$ values. 
Furthermore, \app takes advantage of population archives that not only carry information regarding the highest-performing individuals but also enforce diversity among archive members.

\begin{table*}
  \centering
  \small
  \caption{Statistical comparison of DBS values for different search methods at different values of $t_b$ and $d_\mathit{th}$. Comparisons with no results are specified as $N/A$. Such cases happen when $A$ or $B$ have no samples to compare.}
  \label{tab:rq1-stat-test}
  \resizebox{0.95\textwidth}{!}{
  \begin{tabular}{ccccccccccccccc}
  \toprule
    &\multicolumn{2}{c}{Comparison}&\multicolumn{12}{c}{$DBS$}
    \\
    \cmidrule(lr){2-3}\cmidrule(lr){4-15}
    &\multirow{2}{2em}{$A$}&\multirow{2}{2em}{$B$}&\multicolumn{2}{c}{$t_b=0.01$}&\multicolumn{2}{c}{$t_b=0.03$}&\multicolumn{2}{c}{$t_b=0.05$}&\multicolumn{2}{c}{$t_b=0.10$}&\multicolumn{2}{c}{$t_b=0.15$}&\multicolumn{2}{c}{$t_b=0.20$}
    \\
    \cmidrule(lr){4-5}\cmidrule(lr){6-7}\cmidrule(lr){8-9}\cmidrule(lr){10-11}\cmidrule(lr){12-13}\cmidrule(lr){14-15}
    &&&$p$&$\hat{A}_{AB}$&$p$&$\hat{A}_{AB}$&$p$&$\hat{A}_{AB}$&$p$&$\hat{A}_{AB}$&$p$&$\hat{A}_{AB}$&$p$&$\hat{A}_{AB}$\\

    \midrule
    \multirow{3}{2em}{$d_\mathit{th}=0.1$}&$MLCSHE$&$RS$&$N/A$&$N/A$&$N/A$&$N/A$&$N/A$&$N/A$&$\num{1.46e-04}$&$1.00$&$\num{1.78e-04}$&$1.00$&$\num{1.83e-04}$&$1.00$\\
    &$MLCSHE$&$GA$&$N/A$&$N/A$&$N/A$&$N/A$&$\num{1.00e+00}$&$0.51$&$\num{4.40e-04}$&$0.97$&$\num{1.83e-04}$&$1.00$&$\num{1.83e-04}$&$1.00$\\
    &$RS$&$GA$&$N/A$&$N/A$&$N/A$&$N/A$&$N/A$&$N/A$&$\num{1.46e-04}$&$0.00$&$\num{1.78e-04}$&$0.00$&$\num{1.83e-04}$&$0.00$\\
    \cmidrule(lr){1-3}\cmidrule(lr){4-15}
    \multirow{3}{2em}{$d_\mathit{th}=0.2$}&$MLCSHE$&$RS$&$N/A$&$N/A$&$N/A$&$N/A$&$N/A$&$N/A$&$\num{1.44e-04}$&$1.00$&$\num{1.73e-04}$&$1.00$&$\num{1.77e-04}$&$1.00$\\
    &$MLCSHE$&$GA$&$N/A$&$N/A$&$N/A$&$N/A$&$\num{4.88e-02}$&$0.77$&$\num{1.79e-04}$&$1.00$&$\num{1.81e-04}$&$1.00$&$\num{1.73e-04}$&$1.00$\\
    &$RS$&$GA$&$N/A$&$N/A$&$N/A$&$N/A$&$N/A$&$N/A$&$\num{1.44e-04}$&$0.00$&$\num{1.73e-04}$&$0.00$&$\num{6.32e-03}$&$0.14$\\
    \cmidrule(lr){1-3}\cmidrule(lr){4-15}
    \multirow{3}{2em}{$d_\mathit{th}=0.3$}&$MLCSHE$&$RS$&$N/A$&$N/A$&$N/A$&$N/A$&$N/A$&$N/A$&$\num{1.38e-04}$&$1.00$&$\num{1.63e-04}$&$1.00$&$\num{1.70e-04}$&$1.00$\\
    &$MLCSHE$&$GA$&$N/A$&$N/A$&$N/A$&$N/A$&$\num{3.35e-03}$&$0.89$&$\num{1.74e-04}$&$1.00$&$\num{1.72e-04}$&$1.00$&$\num{1.75e-04}$&$1.00$\\
    &$RS$&$GA$&$N/A$&$N/A$&$N/A$&$N/A$&$N/A$&$N/A$&$\num{1.36e-04}$&$0.00$&$\num{1.62e-04}$&$0.00$&$\num{5.45e-03}$&$0.13$\\
    \bottomrule
\end{tabular}}
\end{table*}

Our visual observations are supported by the results of the statistical comparisons provided in Table~\ref{tab:rq1-stat-test}.
Columns $A$ and $B$ indicate the search methods being compared.
Columns $p$ and $\hat{A}_{AB}$ indicate statistical significance and effect size, respectively, when comparing A and B in terms of \textit{DBS} at different $t_b$ and $d_\mathit{th}$ values.
Comparisons with no results are denoted as N/A; it happens when $A$ or $B$ have no boundary search results to compare.
Given a significance level of $\alpha=0.01$, the differences between \app and other methods are significant when $t_b \geq 0.05$, except when $t_b=0.05$ and $d_\mathit{th}=0.1$---that is when the very low threshold makes it infeasible to find many complete solutions that are both \textit{distinct enough} from each other and \textit{close enough} to the hazard boundary---for which the average \textit{DBS} of \app is only slightly higher than that of GA.
Moreover, $\hat{A}_{AB}$ is always greater than $0.71$ when $A=\text{\MLCSHE}$, indicating that \app always has a large effect size when compared to other search methods. 
Therefore, we conclude that, for $t_b$ values that require practical numbers of simulations to find a sufficient number of distinct boundary solutions for safety monitoring, \app yields better results than GA and RS.

\begin{tcolorbox}
For boundary closeness thresholds that require practical numbers of simulations to find a sufficient number of distinct boundary solutions for safety monitoring, \app is significantly more effective than GA and RS with high effect size, meaning that \app finds significantly more diverse regions near the hazard boundary.
\end{tcolorbox}

\subsection{RQ2: Efficiency}\label{sec:efficiency}
\subsubsection{Methodology}\label{sec:rq2-method}
To answer RQ2, we follow the same methodology as for RQ1, including the hyperparameters and 10 repeats for each method, except for the search (simulation) budget. 
Specifically, we measure \textit{DBS} across different methods while varying the simulation budget from 10\% (130 simulations) to 100\% (1300 simulations) in steps of 10\%. We then report and analyze how the effectiveness values of different methods vary over time. 

\subsubsection{Results}\label{sec:rq2-results}

\begin{figure*}
\centering
\begin{subfigure}{0.75\textwidth}
  \centering
  \includegraphics[width=\linewidth]{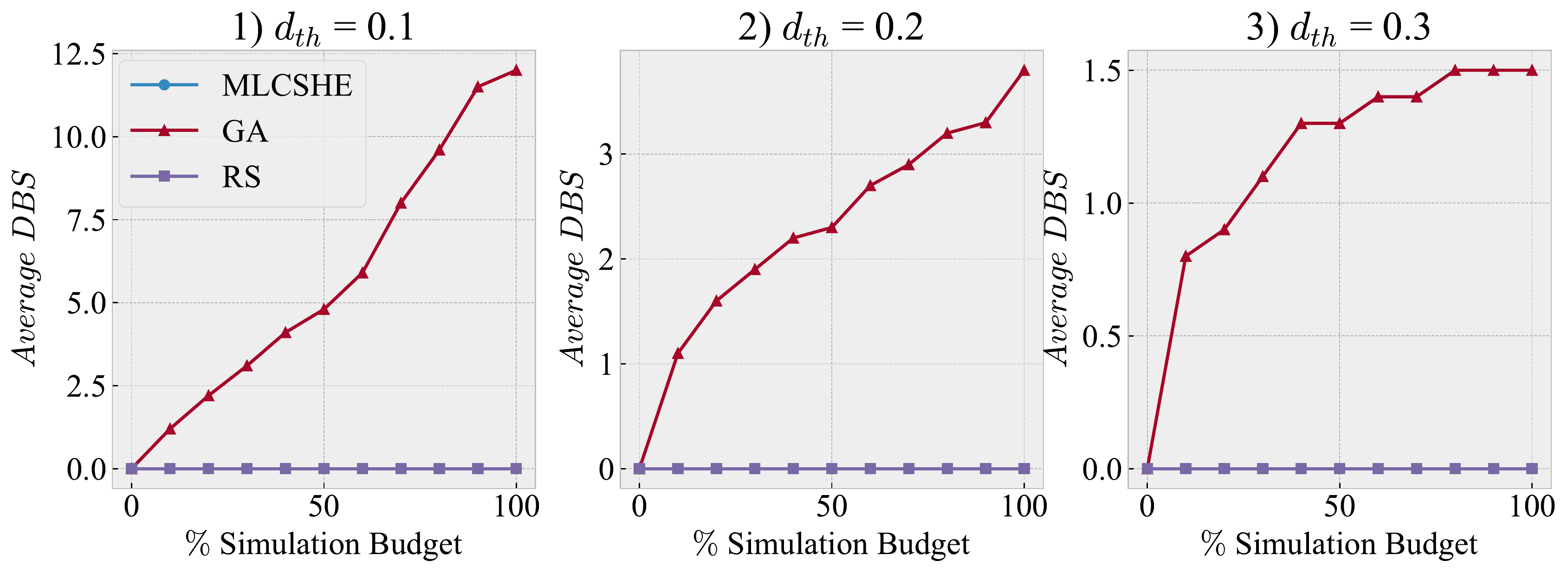}
  \caption{$t_b=0.03$}
  \label{fig:pp-ft-03}
\end{subfigure}\\
\begin{subfigure}{0.75\textwidth}
  \centering
  \includegraphics[width=\linewidth]{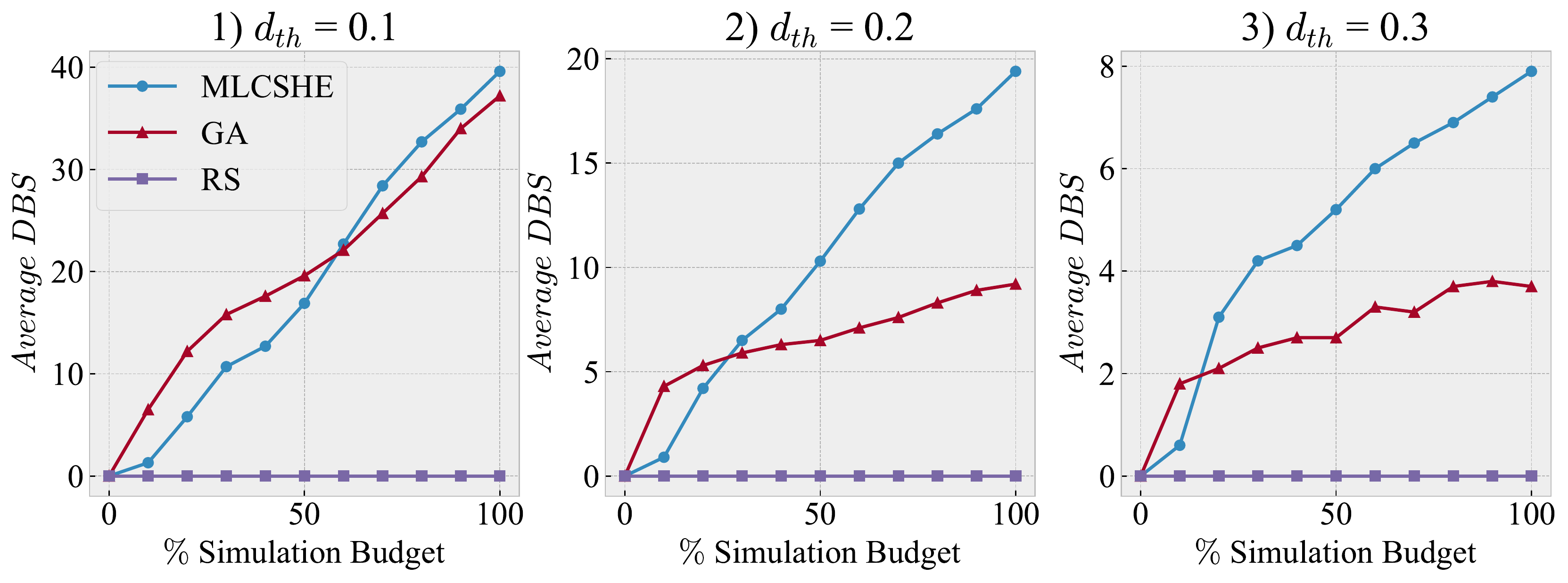}
  \caption{$t_b=0.05$}
  \label{fig:pp-ft-05}
\end{subfigure}\\
\begin{subfigure}{0.75\textwidth}
  \centering
  \includegraphics[width=\linewidth]{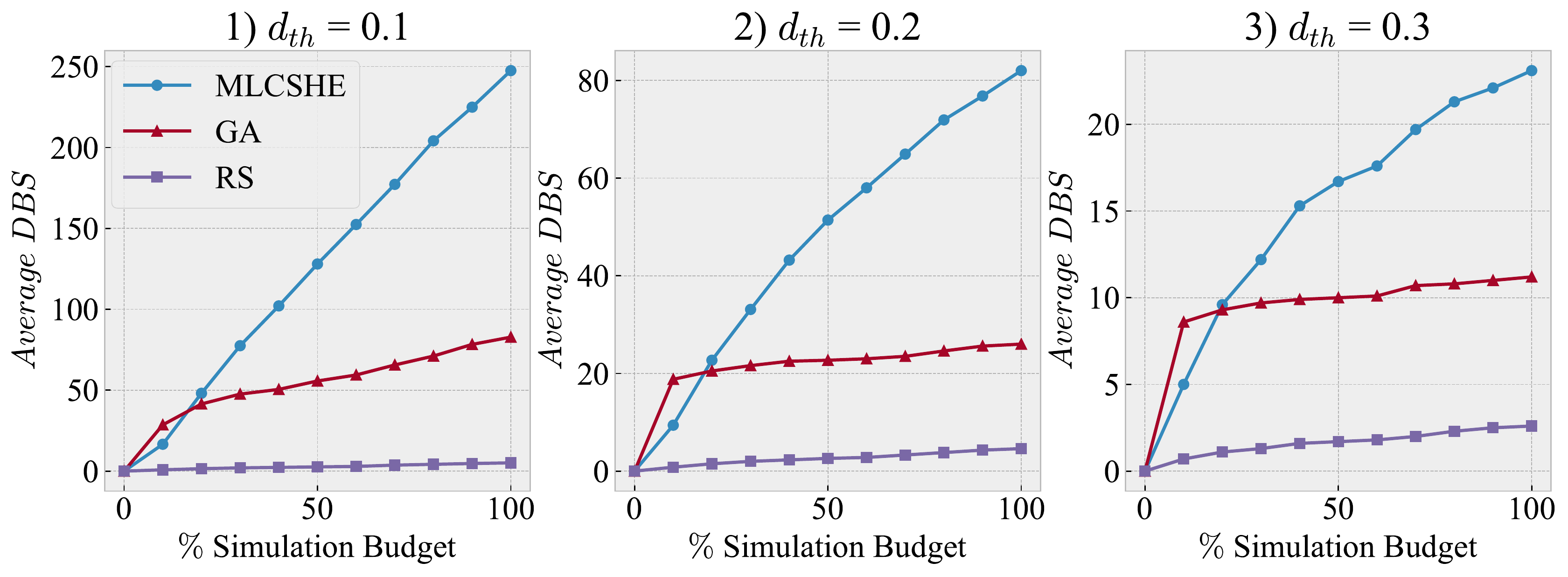}
  \caption{$t_b=0.15$}
  \label{fig:pp-ft-15}
\end{subfigure}
\caption{Plots of \textit{DBS} vs. $\%$ \textit{simulation~budget} for \MLCSHE, GA, and RS, with $d_\mathit{th} \in \{0.1, 0.2, 0.3\}$ and $t_b \in \{0.03, 0.05, 0.15\}$.}
\label{fig:prog-plots}
\end{figure*}

Based on the data we collected in our experiment, we analyzed how all different threshold values for $d_\mathit{th}$ and $t_b$ affect the relationship between the percentage of simulation budget consumed and the average \textit{DBS} values for 10 runs across MLCSHE, GA, and RS. 
In \figurename~\ref{fig:prog-plots}, we selected three $t_b$ values ($0.03$, $0.05$, and $0.15$) that, together, are representative of the overall trends.
The remaining plots\footnote{The plots for $t_b  = 0.1$ and $t_b = 0.2$ are very similar to \figurename~\ref{fig:pp-ft-15}.} are available in the supporting material (see Section~\ref{sec:data-availability}).

On the one hand, \figurename~\ref{fig:pp-ft-03} shows that only GA finds a few boundary solutions when $t_b = 0.03$. Although GA does not reach a plateau for $d_\mathit{th} \leq 0.2$, the numbers of distinct boundary solutions found by GA are not enough for safety monitoring as already discussed in Section~\ref{sec:rq1-results}. 
On the other hand, \figurename~\ref{fig:pp-ft-05} and \figurename~\ref{fig:pp-ft-15} show that
\app leads to significantly higher \textit{DBS} once the consumed budget is above 10\%, except when $t_b = 0.05$ and $d_\mathit{th} = 0.1$ for reasons that we already discussed in Section~\ref{sec:rq1-results}.

We suspect that the results during the first 10\% of the simulation budget can be explained by the initial overhead of \app: since it simulates all possible complete solutions that can be generated by joining the scenario and MLC output populations in the first generation, it could complete only one search generation while GA could complete two or more generations.
However, \app continues to find new distinct complete solutions near the boundary as the budget increases, whereas GA quickly starts to stagnate and reach a plateau.
As a result, after only spending 20\% of the total budget, \app always significantly outperforms GA.

Note that, even though we had to set the maximum simulation budget to 1,300 simulations due to the large size of experiments and the unavoidable limitations in computational resources, the \textit{DBS} values of \app keep increasing until the budget is exhausted for practical boundary closeness thresholds ($t_b \geq 0.05$). This suggests that \app is able to find considerably more boundary solutions with more simulation budget, when available. 

\begin{tcolorbox}
\app is significantly more efficient than GA and RS for practical boundary closeness thresholds: \app finds significantly more diverse regions that overlap with the hazard boundary at a faster rate than GA and RS.
\end{tcolorbox}

\subsection{Discussion}\label{sec:discussion}

\subsubsection{Interpretability of Boundary Region}\label{sec:interpretability}

Since the boundary complete solutions found by \app and other methods are for safety monitoring, one might wonder if we could obtain interesting insights from such boundaries regarding the characteristics leading to high risks of safety violations.

However, the concept of meaningful boundaries is not relevant here as we are not looking for boundaries for a specific MLC implementation but for boundaries that should not be approached by any implementation for given scenarios. 
Incorrect MLC implementations can yield arbitrary outputs that may lead, for certain scenarios, to violations. 
Why that is the case for certain scenarios and not others is extremely difficult to explain as it requires going into the details of how the system uses these outputs across different scenarios.

Nevertheless, the violations identified by \app are the result of executing the system (including the MLC) in interaction with the simulation environment. 
Therefore, they are \textit{real} violations regardless of whether engineers can interpret them.

\subsubsection{Threats to Validity}
\label{sec:threats}

This section discusses potential threats to the validity of our results, namely internal, external, conclusion and construct validity~\cite{val-1,val-2,val-3}.

\textit{Internal Validity.}
Internal validity is concerned with the accuracy of the cause-and-effect relationships established by the experiments.

As mentioned in Section~\ref{sec:rq1-method}, the actual number of executed simulations is slightly higher than the allocated simulation budget (1,300) for the population-based methods (i.e., \MLCSHE and GA). Although the same budget should be used for different methods for a fair comparison, the deviations are so small (less than $5\%$ of the allocated budget) that they cannot significantly impact the results in terms of effectiveness and efficiency.

Another potential threat to internal validity is that the hyperparameter values for GA can affect the results. For example, one might want to intentionally increase the mutation rate of GA to improve the diversity of the complete solutions found by GA. However, it could make GA similar to a Random Search (RS) and could substantially reduce its performance~\cite{Mirjalili2019GA}, which was also confirmed in our preliminary evaluation results. To mitigate this threat, as mentioned in Section~\ref{sec:rq1-method}, we relied on the values recommended by \citet{Mirjalili2019GA}, which are commonly used in the literature.

\textit{External Validity.}
External validity is concerned with the generalizability of the results.

One notable factor to consider is related to the fact that we have relied only on a specific ADS (Pylot) and simulator (Carla). 
However, Carla is a widely used open-source, high-fidelity simulator, and Pylot was the only component-based AV among those high-ranking in the Carla leaderboard~\cite{CarlaLeadeboard} at the time of our evaluation. 
Moreover, running the experiments on Pylot and Carla took more than 75 days of execution, even with parallelization, making it infeasible to consider additional evaluation subjects.
Nonetheless, further studies involving other ML-enabled Autonomous Systems in autonomous driving as well as other domains, such as aerospace, agriculture, and manufacturing, are required.

Also resulting from the high cost of running experiments is the fact that we could not evaluate different design choices for \MLCSHE using the ADS case study, namely hyperparameters, individual fitness assessment, and archive update strategies.
This might impact the generalizability of our results.
To account for this factor, we relied on two widely used benchmark problems which are widely used by the literature, as referred to in Section~\ref{sec:algorithm}, to tune the hyperparameters of \MLCSHE and decide between alternative strategies.

The generalizability of our results is also affected by the fact that a specific ODD, i.e., urban driving, was considered for the evaluation.
Changing the ODD to highway driving, for example, changes the lower and upper bounds of the scenario parameters, as well as the complete solutions that will be discovered close to the hazard boundary.
However, urban driving is one of the most complex driving ODDs where complicated interactions (and safety violations) between many cars and pedestrians can occur, e.g., at an intersection.
Thus, the hazard boundary related to the urban driving ODD is expected to have a more complex shape than a simpler ODD such as highway driving.
Furthermore, additional ODDs could not be considered due to time and resource constraints, as described above.
We encourage

The specific encoding of scenarios and MLC outputs would be another generalizability factor since it determines the search space, which could significantly affect the effectiveness and efficiency of each search method.
However, for the large search space problems that are common in practice, we expect \MLCSHE to fare increasingly better than GA and RS since \MLCSHE is designed to decompose high-dimensional problems into lower-dimensional subproblems.

\textit{Conclusion Validity.}
Conclusion validity is concerned with the conclusions that can be drawn from the collected data and their statistical significance.
The experiments could only be repeated 10 times, which is less than the widely accepted rule-of-thumb of 30 repetitions. However, as mentioned in Section~\ref{sec:eval-subjects}, more than 1,800 hours were consumed to run the experiments with 10 repetitions.
To account for the statistical error associated with the lower number of repetitions, we report every statistical value with its confidence interval.

\textit{Construct validity.}
Construct validity is concerned with the degree to which the measured variables in the study represent the underlying concept being studied.
In our case, the concept of hazard boundary coverage is operationalized by the \textit{DBS} value, which sufficiently captures both concepts of \textit{closeness} to the hazard boundary (via $t_b$) and \textit{coverage} of the hazard boundary in diverse regions (via $d_\mathit{th}$) at the same time.

\subsection{Data Availability}\label{sec:data-availability}
The search algorithms (i.e., \MLCSHE, GA, RS), the parallel simulation execution module, and the postprocess script are all implemented in Python. The replication package, including the aforementioned implementations, the instructions to set up and configure Pylot and CARLA, the detailed descriptions of the initial conditions used in the experiments, and the detailed results, is available at \cite{Replication_Package}.

\section{Conclusion and Future Work}
\label{sec:conclusion}

In this paper, we presented \MLCSHE, a cooperative coevolutionary search algorithm to effectively and efficiently approximate the systemic hazard boundary of a machine learning component embedded in an ML-enabled autonomous system, given a system-level safety requirement. We address the challenge of the high-dimensional search space and expensive high-fidelity simulations by using cooperative coevolutionary search, which decomposes the problem into two smaller subproblems.
We rely on a probabilistic fitness function that guides the search towards the boundary of probabilistic unsafe regions.
We apply the method to an AV case study, where we run large-scale experiments with parallel simulations to evaluate the effectiveness and efficiency of \MLCSHE.
The evaluation results show that, for practical boundary closeness thresholds, \MLCSHE is significantly more effective and efficient than random search and a standard genetic algorithm in identifying diverse boundary regions.

As part of the future work, we plan to apply \MLCSHE to other AVs as well as other ML-enabled autonomous systems in various domains such as agriculture or aerospace.
Furthermore, we plan to use the hazard boundary approximated using \MLCSHE in developing and evaluating safety monitors, and guiding the testing of ML components being integrated in ML-enabled autonomous systems.

\ifCLASSOPTIONcompsoc
  \section*{Acknowledgments}
\else
  \section*{Acknowledgment}
\fi

The authors are very grateful to Auxon Corporation for their financial support and to Zachary Pierce for his insightful feedback. 
This work was also supported  through the Natural Sciences and Research Council of Canada (NSERC) Discovery Grant program, Canada Research Chairs (CRC) program, Mitacs Accelerate program, and Ontario Graduate Scholarship (OGS) program.
Part of the preliminary experiments conducted in this work were enabled by support provided by the Digital Research Alliance of Canada.

\ifCLASSOPTIONcaptionsoff
  \newpage
\fi

\bibliographystyle{IEEEtranN}
\bibliography{IEEEabrv,references}

\end{document}